\documentclass[11pt,a4paper,twoside]{article}

\usepackage{mathrsfs}
\usepackage{enumerate}
\usepackage{amsmath,amssymb}
\usepackage{graphicx,epsfig,natbib}
\usepackage{bibentry}
\usepackage{appendix}
\usepackage{hyperref,color}
\usepackage{caption,subcaption,placeins,float}
\usepackage{tikz,sidecap}
\usepackage{bbm,comment,url}
\usepackage{mathtools}
\usepackage{rotating,multirow}
\usepackage[margin=1in]{geometry}

\usepackage{amsmath,amsfonts,amsthm,bm} 

\usepackage[linesnumbered,ruled]{algorithm2e}
\usepackage{graphicx}

\newcommand{\param}{{\bm{\theta}}}
\newcommand{\hyperp}{{\bm{\psi}}}
\newcommand{\hyperpt}{{\bm{\tilde{\psi}}}}
\newcommand{\Data}{D}
\newcommand{\Npart}{K}
\newcommand{\pap}{r}
\newcommand{\KLE}{E}

\newcommand{\bftheta}{\boldsymbol\theta}
\newcommand{\bfXg}{\textbf{X}_{\boldsymbol\gamma}}
\newcommand{\bfwg}{\textbf{w}_{\boldsymbol\gamma}}
\newcommand{\bfy}{\textbf{y}}

\newcommand{\bfgam}{\boldsymbol\gamma}
\newcommand{\bfSigma}{\boldsymbol\Sigma}

\DeclareMathOperator\Bern{Bern} 
\DeclareMathOperator\Var{Var} 

\title{Improving the identification of antigenic sites in the H1N1 Influenza virus through accounting for the experimental structure in a sparse hierarchical Bayesian model}

\author{Vinny Davies\textsuperscript{1,2,+}, William T. Harvey\textsuperscript{3}, Richard Reeve\textsuperscript{3} and Dirk Husmeier\textsuperscript{2,*} \vspace{0.8cm} \\ {\small \textsuperscript{1}School of Medicine, University of Leeds, Leeds, UK. \textsuperscript{2}School of Mathematics and Statistics,} \\ {\small University of Glasgow, Glasgow, UK. \textsuperscript{3}Boyd Orr Centre for Population and Ecosystem} \\ {\small Health, Institute of Biodiversity, Animal Health and Comparative Medicine,} \\ {\small College of Medical, Veterinary and Life Sciences, University of Glasgow, Glasgow, UK} \\ {\small \textsuperscript{+}V.Davies@Leeds.ac.uk \textsuperscript{*}Dirk.Husmeier@glasgow.ac.uk}}
\date{\vspace{-3ex}}

\begin{document}
\maketitle

\begin{abstract}
Understanding how genetic changes allow emerging virus strains to escape the protection afforded by vaccination is vital for the maintenance of effective vaccines. In the current work, we use structural and phylogenetic differences between pairs of virus strains to identify important antigenic sites on the surface of the influenza A(H1N1) virus through the prediction of haemagglutination inhibition (HI) assay, pairwise measures of the antigenic similarity of virus strains. We propose a sparse hierarchical Bayesian model that can deal with the pairwise structure and inherent experimental variability in the H1N1 data through the introduction of latent variables. The latent variables represent the underlying HI assay measurement of any given pair of virus strains and help account for the fact that for any HI assay measurement between the same pair of virus strains, the difference in the viral sequence remains the same. Through accurately representing the structure of the H1N1 data, the model is able to select virus sites which are antigenic, while its latent structure achieves the computational efficiency required to deal with large virus sequence data, as typically available for the influenza virus. In addition to the latent variable model, we also propose a new method, block integrated Widely Applicable Information Criterion (biWAIC), for selecting between competing models. We show how this allows us to effectively select the random effects when used with the proposed model and apply both methods to an A(H1N1) dataset. 
\end{abstract}

\section{Introduction}

Human influenza viruses are a major cause of morbidity and mortality worldwide, with seasonal epidemics of influenza estimated to result in 3-5 million cases of severe illness and 250,000-500,000 deaths \citep{WHO09}. Individuals usually mount an effective antibody-mediated immune response following infection or vaccination that provides long-lasting protection against a particular strain of the influenza virus. However, seasonal influenza viruses evolve rapidly and changes to the parts of the virus (termed antigens) recognised by the immune system enable the virus population to evade existing immunity and individuals experience recurrent infections. Furthermore the effectiveness of the vaccine, which remains the most effective means of disease prevention, depends on the constituents being well matched to circulating viruses. The continuing antigenic evolution of influenza viruses requires a World Health Organization (WHO) coordinated Global Influenza Surveillance and Response System (GISRS), responsible for the identification of new genetic and antigenic variants among circulating viruses in order to ensure that influenza vaccine components reflect the antigenic characteristics of circulating viruses \citep{WHO09}.

Influenza viruses are classified into three distinct types (A, B and C), of which A and B viruses circulate globally in humans and are responsible for seasonal epidemics. Influenza A viruses are particularly diverse and are further classified into subtypes (e.g. A(H1N1)). The influenza vaccine comprises strains of A(H1N1), A(H3N2) and B viruses predicted to elicit the most effective immune responses against circulating viruses in the forthcoming influenza season  \citep{Barr14}. Vaccination provides minimal protection across subtypes and effectiveness within subtype is maximised when the vaccine virus is more antigenically similar to circulating viruses. Genetic mutations cause amino acid substitutions in the surface proteins of the influenza virus that affect recognition by the human immune system. The ever-changing antigenic characteristics of influenza viruses require that the vaccine formulation is reviewed twice annually and is frequently updated to maintain protection.

The motivation behind this work is to develop models that predict antigenically significant amino acid residues within the influenza surface proteins. An improved understanding of the genetic basis of antigenic evolution has the potential to aid the vaccine selection process in a variety of ways. The development of \emph{in silico} models which can predict both antigenic residues and the likely cross-protection offered by candidate vaccine viruses strains is vital for directing these experiments in an efficient manner and reducing the amount of experimental work that must be carried out. In addition to the identification of emerging antigenic variants, experts must anticipate which viruses are likely to predominate in forthcoming epidemic seasons. Models that improve our knowledge of the contributions of changes to amino acid residues to antigenic evolution have the capacity to enhance the existing evolutionary models currently used to predict which strains will increase or decrease in frequency through time (e.g. \cite{Luksza14}).

In order to infer the antigenic importance of genetic changes that have occurred during the evolution of the virus we require both genetic data and a measure of antigenic similarity. Antigenic properties of influenza viruses are largely determined by the surface protein haemagglutinin (HA), which agglutinates red blood cells during infection, causing them to clump together. Human antibodies recognise exposed parts of the HA, and bind to it, inhibiting the agglutination. Amino acid substitutions (changes) on the surface of the HA protein cause loss of recognition by human antibodies, and the haemagglutination inhibition (HI) assay, which determines the extent to which antibodies inhibit agglutination of red blood cells, is commonly used for antigenic characterisation of circulating viruses \citep{Hirst42,WHO11}. The HI assay is used to assess the antigenic similarity of a circulating test virus to each of a panel of reference strains that typically include the current vaccine strain and a range of potential future vaccines. It approximates the degree of protection each reference strain would provide against the test virus by recording the maximum dilution at which antibodies in a sample of antiserum from a ferret exposed to a particular reference strain remain able to inhibit the clumping of red blood cells (haemagglutination) by a sample of the test virus. A high titre in the test corresponds to a high dilution of the antiserum and therefore a low concentration of the antiserum being sufficient to cause inhibition; a low titre conversely corresponds to a low dilution and a high concentration of the antiserum being required. An antiserum is therefore typically able to inhibit the virus used to produce the antiserum at high dilutions, but higher concentrations (and hence lower titres) are required to inhibit test viruses that are antigenically more dissimilar. Higher HI titres indicate antigenic similarity, and HI titres typically decrease with increasing genetic distance between reference and test viruses.

Each HI titre can be associated with genetic data relating to differences between the reference and test viruses used in the assay. The contributions of individual amino acid substitutions to antigenic evolution can be predicted by comparing amino acid sequences of reference and test viruses. Furthermore, terms that describe the evolutionary relatedness of viruses can also be used to explain differences in HI titres. In addition to antigenic similarity, HI titres also reflect variation in the binding strength of both antiserum and test virus. Variation in each of these binding strengths can also be modelled using evolutionary terms. As the experiments used to determine the HI titre test the difference between pairs of reference and test viruses, HI titres will have the same corresponding explanatory variables related to the amino acid substitutions regardless of which virus is used as the reference virus and which the test virus. This is not, however, the case for the variables describing the evolutionary relatedness of the viruses as the variables related to the binding strengths of antiserum and test virus depend on which of the pair was used in each context. HI titres are also affected by experimental variability caused by for example differences in the dilution of reagents between experiments carried out on different dates. The structure of the dataset will be further explained in Section~\ref{sec:data2}.

Various methods have been proposed to account for the experimental variation in the measurements and select the variables which cause the changes in the measured antigenic variability. Originally \cite{Reeve10} used mixed-effects models, e.g. \cite{PinheiroBates00}, to predict the antigenic similarity of foot-and-mouth disease virus (FMDV) strains. The authors first selected the random effect components and then added terms to account for the evolutionary history of the viruses. Finally a univariate test for significance was used on the residue variables, with a p-value of less than 0.05 corresponding to an antigenically important residue. A similar method has also been applied by \cite{Harvey16} to the influenza A(H1N1), using versions of the datasets used here.

\cite{Davies14} then introduced a sparse hierarchical Bayesian model for detecting relevant antigenic sites in virus evolution (SABRE) and showed how it outperformed the method of \cite{Reeve10}. The SABRE model uses spike and slab priors, as proposed in \cite{MitchellBeauchamp88}, to improve variables selection and outperform the mixed-effects Least Absolute Shrinkage and Selection Operator (LASSO) \citep{Tibshirani96,Schelldorfer11}. In the SABRE model, the spike and slab priors are integrated into a Bayesian hierarchical mixed-effects model, allowing for consistent inference of all parameters and hyper-parameters, and inference that borrows strength by the systematic sharing and combination of information; see \cite{Gelman13}. \cite{Davies17_COST} improved the SABRE model through the addition of a biologically significant intercept parameter and increased conjugacy between parameter.

The SABRE models of \cite{Davies14,Davies17_COST} do not however fully take into account the structure of the data and are not computationally efficient enough to work with the H1N1 dataset. The structure of the data comes from the fact that the HI assay is often repeated multiple times for the same reference and test virus pair. Correspondingly, the genetic and evolutionary data will be the same for any two measurements where the same reference and test viruses are used. However as the full set of explanatory variables explicitly depend on which of the two viruses are used as the reference virus and which was used as the test virus, it is worth noting that a given pair of viruses will give different explanatory variables if the strains used as reference and test virus are switched. We can use the described structure to improve the accuracy of the SABRE model and increase its computational efficiency such that it can now be used on the H1N1 dataset. In the current work we introduce an extended version of the SABRE model, the eSABRE model, through the use of a latent variable model which better matches the structure of the data. More precisely we introduce latent variables to represent the underlying HI assay of any given pair of reference and test virus.

In addition to selecting the fixed-effects, it is also important to choose the random effect components. To do the selection we introduce a variation of the Widely Applicable Information Criterion (WAIC) \citep{Watanabe10}, block integrated WAIC (biWAIC) based on integrated WAIC (iWAIC) as proposed in \cite{Li15}. biWAIC takes into account the specific structure of the eSABRE model and integrates over the latent variables. We describe how this converges to a particular form of Cross Validation (CV) and use a simulation study to quantify the improvement it offers over non-integrated WAIC (nWAIC).

In this paper we evaluate the advantages of the eSABRE model over the previously proposed conjugate SABRE model. We use simulated datasets that mimic the structure of the H1N1 dataset to show how it offers an improvement in variable selection, as well as an increase in computational efficiency. We also propose and test biWAIC on the simulated datasets to quantify its improvement in selecting random effect components within the eSABRE model. Finally we apply biWAIC with the eSABRE model to the H1N1 dataset and identify a number of known and potential antigenic sites, comparing the results with those of \cite{Harvey16}.

\section{Data}
\label{sec:data2}	

The antigenic data analysed comprised pairwise measures of antigenic similarity of viruses of the A(H1N1) subtype obtained using the haemagglutination inhibition (HI) assay. In these experiments, antiserum created by exposing a ferret to a particular reference virus is measured in terms of its ability to inhibit the binding of red blood cells (haemagglutination) by a sample of a second virus, the test virus. Within the dataset, there are multiple measurements, $\textbf{y}$, which are taken from the same pair of reference and test virus, $p$, but they are often carried out under different experimental conditions. In the case of two measurements where the same pair of reference and test virus was used, $y_1$ and $y_2$, the experimental conditions, $\textbf{Z}$, for these observations can vary, i.e. $\textbf{Z}_1 \neq \textbf{Z}_2$ or $\textbf{Z}_1 = \textbf{Z}_2$. However the corresponding explanatory variables, $\textbf{X}$, will remain the same, $\textbf{X}_1 = \textbf{X}_2$, whenever the same pair of reference and test viruses is used.

For each observation, $y_i$, the explanatory variables, $\textbf{X}_i$, include variables that give the differences in protein structure and evolutionary history between the reference and test viruses. As an individual strain will always have the same protein structure, for any pair of virus strains the differences in protein structure remain identical whenever the experiments are carried out, regardless of which strain is used as the reference strain. More precisely, the explanatory variables, $\textbf{X}$, that give the differences in the protein structure look at whether there is a presence (1) or absence (0) of an amino acid substitution at each specific residue which is exposed on the surface of haemagglutinin (HA) protein. Not all of these amino acid substitutions affect antigenicity but any important changes are likely to result in antigenic differences and a reduction in the observed HI assay measurements, $\textbf{y}$.

The evolutionary or phylogenetic tree representing the relatedness of studied viruses was constructed as described by \cite{Harvey16}. While the tree remains constant in our analyses, the evolutionary variables associated with the structure of the tree may change according to whether a given strain is being used as a reference or test virus in the assay; see Section~6 of the supplementary materials of \cite{Davies17_COST}. When we are unable to attribute antigenic differences to amino acid changes directly, it may be possible to attribute the variation to an explanatory variable representing a point in the evolution of the virus (specifically a branch of the phylogenetic tree) where the antigenic properties of the virus have changed. For any two viruses in the evolutionary tree, a path can be traced through the tree along the branches that separate them. For each observation, $y_i$, the evolutionary variables associated with antigenic change look at whether a branch does (1) or does not (0) form part of the path through the evolutionary tree separating the reference and test virus used in the HI assay.

In addition to measuring antigenic similarity, HI titres are affected by the binding strength of both antiserum and test virus. Variables were added to account for the different effects that evolution has on antiserum and test virus binding strength. In addition to the evolutionary variables related to antigenic change described above, it is possible to include evolutionary variables associated with evolution in the immunogenicity or avidity of the virus \citep{Davies17_COST}. As these variables explicitly depend on which of the two viruses in any given pair was used to create the antiserum and which was present in the assay as a test virus sample, they are different for a given pair of viruses depending on which of the strains is used as the reference virus and which the test virus. Variation in immunogenicity results in antiserum from different strains that vary in their baseline titres, while variation in avidity results in test viruses that tend to have systematically higher or lower titres irrespective of which reference strain they are being tested against. Full details of the evolutionary variables are given in Section~6 of the supplementary materials of \cite{Davies17_COST}.

HI assay measurements usually contain significant experimental variation and it is therefore necessary to include random effects. For the A(H1N1) dataset the possible random effects are lab conditions, reference virus and test virus. Lab conditions account for differences in the experimental conditions seen on particular days such as the dilution of reagents. The reference and test virus effects account for antiserum and viruses that tend to have systematically higher or lower HI titres in all assays in which they are used.

\subsection{Influenza A(H1N1)}
\label{sec:H1N1}

Influenza A(H1N1) viruses re-entered the human population in 1977 and co-circulated with viruses of a second influenza A subtype, A(H3N2), and influenza B viruses until their replacement by a novel, distantly related lineage of A(H1N1) viruses in the 2009 swine-origin pandemic \citep{Barr14}. During the period 1977--2009, the influenza vaccine included an A(H1N1) strain which had to be updated on nine occasions in order to remain antigenically matched to, and therefore capable of protecting the human population from, circulating strains. The dataset analysed here comprises 43 A(H1N1) viruses collected from 1978 to 2009 that were each used as both as reference strains contributing antiserum to the HI assay and as test viruses. From these viruses, 570 different reference and test virus pairs, $p$, were tested resulting in 15,693 HI assay measurements, $\bfy$. Once residues with incomplete genetic data were removed, there were 275 explanatory variables consisting of 53 surface exposed residues and 226 variables related to the phylogenetic data.

\begin{figure*}[t]
				\centerline{\includegraphics[width=0.5\textwidth]{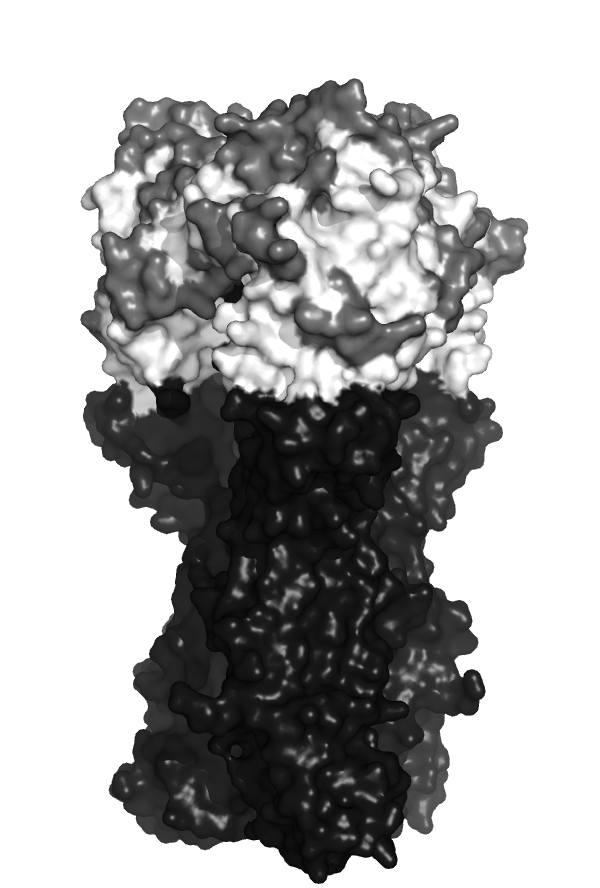}}
				\caption{Three-dimensional structure of the influenza A(H1N1) haemagglutinin (HA) protein coloured by antigenic status. HA is exposed on the virus surface, is responsible for binding to host cells and is the primary target for the host immune system. Known antigenic sites and the receptor-binding site where changes are also expected to cause variation in the HI assay are shown in dark grey (proven regions). Plausible antigenic regions in the head domain of HA are shown in light grey. Implausible antigenic regions in the stalk domain are shown in black, as are surface-exposed areas of the HA2 part of the protein which was not included in our analysis. This model representation of the surface of HA is based on the resolved structure of influenza A(H1N1) strain A/Puerto Rico/8/34 \citep{Gamblin04}.}
				\label{fig:CapsidClassification}
\end{figure*}

For influenza viruses, the haemagglutinin (HA) surface protein is responsible for binding to host cells and is also the major target for neutralising antibodies \citep{SkehelWiley00}. Consequently changes to the HA structure are usually responsible for the requirement to update vaccine components. The structure of HA given in Figure~\ref{fig:CapsidClassification} can be broadly divided into the stalk domain which connects to the virus particle and a head domain which contains the residues involved in binding to the host cell. Experimental studies have identified that the major antigenic regions of HA are protruding areas in the head of the HA protein surrounding the receptor-binding site \citep{SkehelWiley00}. For A(H1N1), these experiments have identified four antigenic sites \citep{Caton82}, however other residues are also known to be important \citep{McDonald07}. We classify residues as proven if they belong to any of the four antigenic sites or have other experimental support for their role in antigenicity (e.g. \citet{McDonald07}), or if they belong to the receptor-binding site where substitutions are expected to influence HI titres via changes to virus receptor-binding strength. These residues are shown in dark grey in Figure~\ref{fig:CapsidClassification}. Other HA residues that are exposed on the surface of the head domain are considered to be plausible antigenic residues, while residues belonging to the stalk domain are considered unlikely to play a role in antigenic change and are therefore considered implausible.  Plausible and implausible antigenic candidate residues are shown in light grey and black respectively in Figure~\ref{fig:CapsidClassification}.

\section{The eSABRE Model}
\label{sec:eSABRE}

The eSABRE model is based on the conjugate SABRE model described in \cite{Davies17_COST} but with a likelihood that better takes into account the data structure. The change in the structure is given in Section~\ref{sec:M2_Likelihood} with the remaining sections defining the prior distributions of the eSABRE model, keeping to those used for the conjugate SABRE model as close as possible. Finally, the model is shown as a Probabilistic Graphical Model (PGM) in Figure~\ref{fig:extDAG} and the parameters are sampled from the posterior distribution using Markov chain Monte Carlo (MCMC) described in Section~\ref{sec:post_inf}.

\subsection{Latent Variable Based Likelihood}
\label{sec:M2_Likelihood}

Previously \cite{Davies17_COST} used the following likelihood similar to classical mixed-effects models:
\begin{align}
	\label{eq:SABRE_like}
	\textbf{y} \sim \mathcal{N}(\textbf{y} | \textbf{1}w_0 + \bfXg \bfwg + \textbf{Z}\textbf{b}, \sigma_\varepsilon^2 \textbf{I}).
\end{align}
In \eqref{eq:SABRE_like}, the response, log HI assay, is given by $\textbf{y} = (y_1,\dots,y_N)^\top$, where $N$ is the number of responses. The random-effects design matrix, $\textbf{Z}$, is set to be a the matrix of indicators with $N$ rows and $||\textbf{b}||$ columns, where $||.||$ indicates the length of the vector and $\textbf{b}$ is a column vector of random-effect coefficients. The explanatory variables, $\textbf{X}$, are given as a matrix of $J+1$ columns and $N$ rows, where $J$ is the number of explanatory variables. The explanatory variables contain indicators of amino acid changes at different residues or information on the phylogenetic structure where the first column is a column full of ones for the intercept. Of the explanatory variables, $\textbf{X}$, only the variables which are relevant to the prediction of $\textbf{y}$, $\bfXg$, are included in \eqref{eq:SABRE_like} dependant on $\bfgam = (\gamma_1, \ldots, \gamma_J)^\top \in \{0,1\}^J$. The relevance of the $j$th column of $\textbf{X}$ is determined by $\gamma_j \in \{0,1\}$, where feature $j$ is said to be relevant if $\gamma_j=1$. Similarly $\bfwg$ is given as the column vector of regressors, where the inclusion of each parameter is dependent on $\bfgam$.

\eqref{eq:SABRE_like} gives a general model which can be used in a variety of different contexts, it does not, however, completely account for the structure of the data used to model antigenic variability and described in Section~\ref{sec:data2}. While the experimental conditions, represented by the random effects, usually vary, each pair of reference and test viruses will have the same explanatory variables. As a result we can introduce latent variables, $\boldsymbol\mu_{\textbf{y}}$, into the model, where each $\mu_{\textbf{y},p}$ represents the unknown true value of the HI assay of any given pair of reference and test viruses, $p$.

\begin{figure*}[t]
				\centerline{\includegraphics[width=\textwidth]{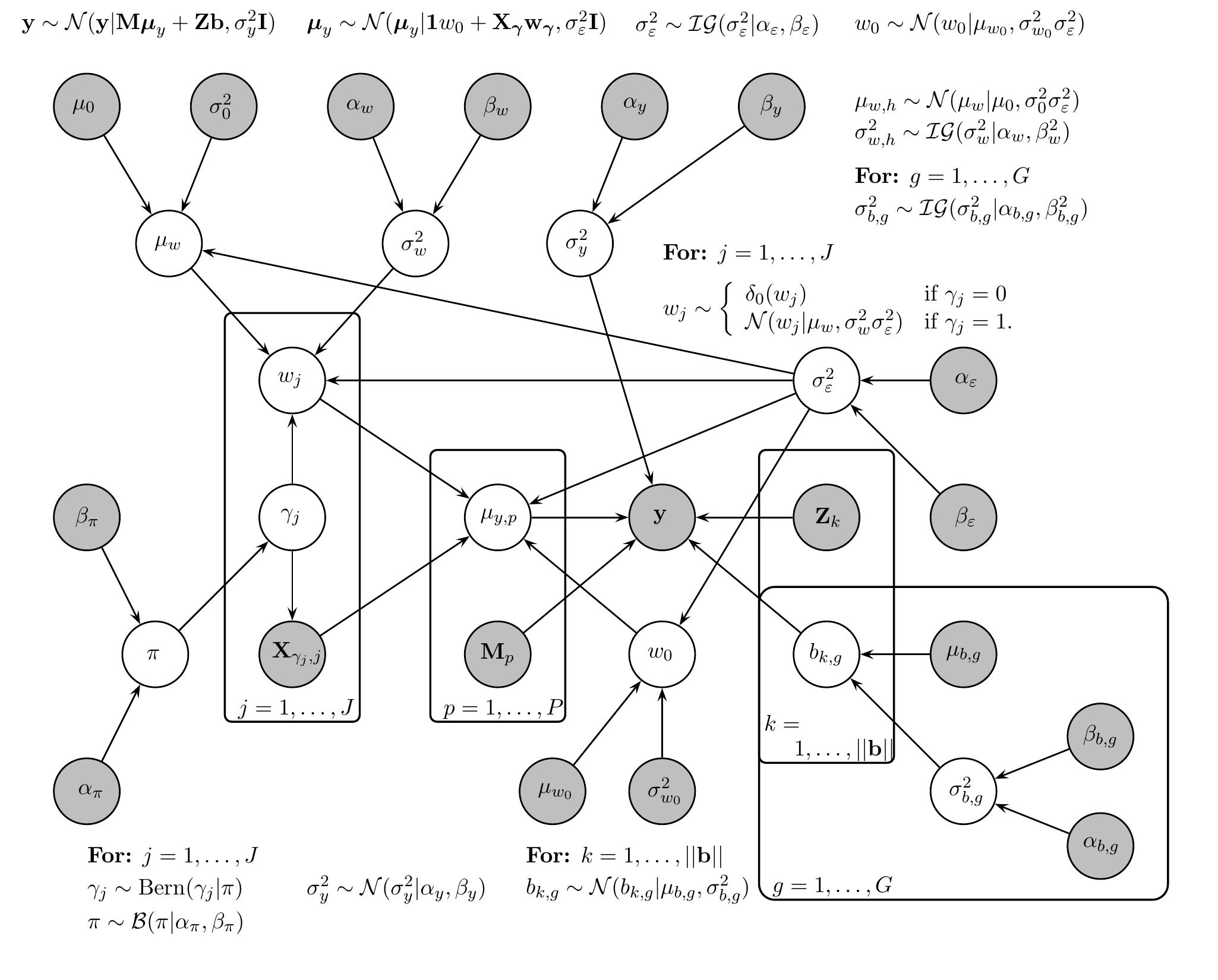}}
				\vspace{-0.7cm}
				\caption{Compact representation of the eSABRE model as a PGM. The \emph{grey} circles refer to the data and fixed (higher-order) hyperparameters, while the \emph{white} circles refer to parameters and hyperparameters that are inferred.}
				\label{fig:extDAG}
\end{figure*}

The introduction of the latent variables, $\boldsymbol\mu_{\textbf{y}}$, into the model results in the following distribution for $\bfy$:
\begin{align}
	\label{eq:like_lat1}
	\textbf{y} &\sim \mathcal{N}(\textbf{y} | \textbf{M} \boldsymbol\mu_{\textbf{y}} + \textbf{Z} \textbf{b}, \sigma_y^2 \textbf{I})
\end{align}
where $\textbf{M}$ is a design matrix which ensures that each $y_i$ has the latent variable, $\mu_{\textbf{y},p}$, which corresponds to its given pair of reference and test viruses, $p$. The random effects are added into the likelihood as some of these factors, e.g. date, affect measurements at the individual level, i.e. they are different for each $y_i$.

We then wish to infer the values of the HI assay measurements of the pairs of virus strains, $\boldsymbol\mu_{\textbf{y}}$, based on the differences in the protein structure and evolutionary history of the virus:
\begin{align}
	\label{eq:like_lat2}
	\boldsymbol\mu_{\textbf{y}} &\sim \mathcal{N}(\boldsymbol\mu_{\textbf{y}} | \textbf{1}w_0 + \bfXg \bfwg, \sigma_\varepsilon^2 \textbf{I}).
\end{align}
As with the SABRE model, we only wish to use the relevant explanatory variables, $\bfXg$, and corresponding regression coefficients, $\bfwg$. We also include an intercept parameter, $w_0$, as we expect high underlying HI assay measurements when the two virus strains used are the same, i.e. the explanatory variables are equal to zero. The full model is given graphically in Figure~\ref{fig:extDAG}.

The structure given by the two main probability distributions of the eSABRE model, given in \eqref{eq:like_lat1} and \eqref{eq:like_lat2}, has two major advantages over the main probability distribution of the conjugate SABRE model, given in \eqref{eq:SABRE_like}. Firstly it allows us to better attribute the error to the correct part of the model. In the HI assay measurements some of the error comes from variability within the experiments, e.g. getting multiple different results for the same pair of reference and test viruses under the same experimental conditions, and this is accounted for by $\sigma_y^2$. Other error will come from the model fit, e.g. our model not truly replicating the underlying biological process, and this is given by $\sigma_\varepsilon^2$. Improving the attribution of error means our model matches better with the data collection technique and should lead to more accurate results and an improvement in the identification of antigenic sites.

The second advantage of the eSABRE model is significantly improved computational efficiency. For example, to run the MCMC simulations to train the model, as discussed in Section~\ref{sec:post_inf}, it would take the SABRE model weeks or months to sample the required number of iterations to achieve convergence and a reasonable sample size after burn-in. In contrast, with the proposed eSABRE model we are able to achieve these results in a few days. A detailed comparison  will be provided in Section~\ref{sec:M2_SimulationStudies}, Table~\ref{tab:M2_AUROC}. The improvement is a result of reducing the computation required to calculate the conditional posterior distribution of $\bfgam$. In essence, through the introduction of latent variables the eSABRE model reduces the posterior distribution of $\bfgam$ to a multivariate Gaussian distribution of dimension $||\boldsymbol\mu_{\textbf{y}}||$, $||\boldsymbol\mu_{\textbf{y}}||=570$ in the H1N1 dataset, as opposed to dimension $||\bfy||$, $||\bfy||=15,693$, in the SABRE model. This will be discussed in further detail in Section~\ref{sec:sampling}.

\subsection{Noise and Intercept Priors}

The conditional variance of the residuals, given the latent variables, is defined as $\sigma_y^2$ and represents the variance in the error seen in repeated measurements from the HI assay experiments. We give $\sigma_y^2$ the following conjugate prior:
\begin{align}
	\label{eq:sigsq_y}
	\sigma_y^2 \sim \mathcal{IG}(\sigma_y^2 | \alpha_y, \beta_y)
\end{align}
where the hyper-parameters $\alpha_y$ and $\beta_y$ are fixed, as indicated by the grey nodes in Figure~\ref{fig:extDAG}.

The variance of the error in the model fit between the fixed effects and the inferred unknown true values of the HI assay for any given pair of reference and test viruses is represented by $\sigma_\varepsilon^2$ and is given the following prior:
\begin{align}
	\label{eq:sigsq_e_prior}
	\sigma_{\varepsilon}^2 \sim \mathcal{IG}(\sigma_{\varepsilon}^2 | \alpha_\varepsilon, \beta_\varepsilon)
\end{align}
where the hyper-parameters $\alpha_\varepsilon$ and $\beta_\varepsilon$ are fixed. $\sigma_\varepsilon^2$ represents the discrepancy between the unknown true HI assay values for each pair and what is inferred by the fixed effects. $\sigma_\varepsilon^2$ is also included in the distributions for $w_0$, $\bfwg$ and $\mu_w$ (defined in Section~\ref{sec:spike_and_slab}) making the model conjugate rather than semi-conjugate, as discussed in Chapter~3 of \cite{Gelman13}. The advantage of this information sharing is that the error variance in terms of model fit is reflected in the distribution of the regression coefficients and this has been further explored in \cite{Davies17_COST}.

Additionally we also require a prior on our intercept:
\begin{align}
		w_0 \sim \mathcal{N}(w_0 | \mu_{w_0}, \sigma^2_{w_0} \sigma_{\varepsilon}^2 ).
		\label{eq:prior_intercept}
\end{align}
We treat the intercept differently from the remaining regressors, wishing to use vague prior settings so as not to penalise this term and effectively make the model scale invariant \citep{Hastie09}.

\subsection{Spike and Slab Priors}
\label{sec:spike_and_slab}

Spike and slab priors have been used in a number of different contexts and have been shown to outperform $\ell_1$ methods both in terms of variable selection and out-of-sample predictive performance \citep{Mohamed12,Davies14,Davies17_COST}. They were originally proposed by \cite{MitchellBeauchamp88} as a mixture of a Gaussian distribution and Dirac delta spike, but have also been used as a mixture of two Gaussians \citep{GeorgeMcCulloch93,GeorgeMcCulloch97} and as binary mask models, e.g. \cite{Jow14}.

The idea behind the spike and slab prior is that the prior reflects whether the feature is relevant based on the values of $\bfgam$. In this way we expect that $w_j = 0$ if $\gamma_j = 0$, i.e. the feature is irrelevant, and conversely it should be non-zero if the variable is relevant, $w_j \neq 0$ if $\gamma_j = 1$. A conjugate Gaussian prior, with $\sigma_\varepsilon^2$ included for further conjugacy, is then assigned where the feature is relevant and a Dirac delta spike at zero where it is not:
\begin{align}
	\label{eq:w_gam_prior}
	w_j \sim \left\{
	\begin{array}{l l}      
    \delta_0(w_j) & \mbox{if} ~ \gamma_j = 0 \\
    \mathcal{N}(w_j | \mu_w, \sigma_w^2 \sigma_{\varepsilon}^2) & \mbox{if} ~ \gamma_j = 1
	\end{array}\right.
\end{align}
for $j \in 1,\ldots,J$ and where $\delta_0$ is the delta function. Here we have a spike at 0 and as $\sigma_w^2 \sigma_{\varepsilon}^2 \rightarrow \infty$ the distribution, $p(w_j | \gamma_j=1)$, approaches a uniform distribution, a slab of constant height. The prior for the variance of the parameter is then given by:
\begin{align}
	\label{eq:sigsq_w_prior}
	\sigma_w^2 \sim \mathcal{IG}(\sigma_w^2 | \alpha_w, \beta_w).
\end{align}
where $\alpha_w$ and $\beta_w$ are fixed; see Figure~\ref{fig:extDAG}.

In addition to $\sigma_w^2$, we use the hyper-parameter $\mu_w$ to reflect a non-zero prior mean of the regression coefficients, $\bfwg$:
\begin{align}
	\label{eq:mu_w_prior}    
	\mu_w \sim \mathcal{N}(\mu_w | \mu_0, \sigma_0^2 \sigma_{\varepsilon}^2 )
\end{align}
where the hyper-parameters $\mu_0$ and $\sigma_0^2$ are fixed and $\sigma_\varepsilon^2$ is again included in the variance for further conjugacy. This specification comes from our biological understanding of the problem. In the H1N1 dataset we are likely to observe large assay values when the reference and test viruses are the same, represented by the intercept $w_0$. Smaller assay values will then be seen when the reference and test viruses are different, reflecting the fact that any amino acid changes are likely to reduce the similarity between virus strains and meaning that the regression coefficients, $w_j$, are likely to be negative.

The final part of the spike and slab prior is to set a prior for $\bfgam$, the hyperparameters which determine the relevance of the variables:
\begin{align}
	\label{eq:gamma_prior}
	p(\bfgam|\pi) &= {\prod_{j=1}^J} \Bern(\gamma_j|\pi)
\end{align}
where $\pi$ is the probability of the individual variable being relevant. The value of $\pi$ can either be set as a fixed hyper-parameter as in \cite{SabattiJames05}, where the authors argue that it should be determined by underlying knowledge of the problem. Alternatively it can be given a conjugate Beta prior:
\begin{align}
	\label{eq:pi_prior}
	\pi \sim \mathcal{B}(\pi | \alpha_\pi,\beta_\pi)
\end{align}
as has been used here. This is a more general model, which subsumes a fixed $\pi$ as a limiting case for $\alpha_\pi \beta_\pi/((\alpha_\pi + \beta_\pi)^2(\alpha_\pi + \beta_\pi + 1)) \rightarrow 0$ and has also been shown to act as a multiplicity correction in \cite{ScottBerger10}.

\subsection{Random-Effects Priors}
\label{sec:REpriors}

In mixed-effects models the random effects, $b_{k,g}$, are usually given group dependant Gaussian priors where there are $K$ random effects and the group $g$ is defined by $k$, i.e. $b_{k,g}$ is shorthand for $b_{k,g_k}$:
\begin{align}
	\label{eq:b_prior}
	b_{k,g} \sim \mathcal{N}(b_{k,g} | \mu_{b,g},\sigma_{b,g}^2).
\end{align} 
We define this to have a fixed mean, $\mu_{b,g}=0$, and a common variance parameter, $\sigma_{b,g}^2$, with a conjugate Inverse-Gamma prior for each random-effects group $g$:
\begin{align}
	\label{eq:sigsq_b_prior}    
	\sigma_{b,g}^2 \sim \mathcal{IG}(\sigma_{b,g}^2 | \alpha_{b,g}, \beta_{b,g})
\end{align}
where $\alpha_{b,g}$ and $\beta_{b,g}$ are fixed hyper-parameters for each $g$ and we define $\textbf{b} \sim \mathcal{N}(\textbf{b} | \textbf{0}, \boldsymbol\Sigma_{\textbf{b}})$ where $\boldsymbol\Sigma_{\textbf{b}} = diag(\boldsymbol\sigma_{\textbf{b}}^2)$ with $\boldsymbol\sigma_{\textbf{b}}^2 = (\sigma_{b,1}^2,\ldots,\sigma_{b,1}^2,\sigma_{b,2}^2,\ldots,\sigma_{b,G}^2)^\top$ such that each $\sigma_{b,g}^2$ is repeated with length $||\textbf{b}_g||$ as shown in Figure~\ref{fig:extDAG}. We are aware that the application of conjugate Inverse-Gamma priors has been disputed by \cite{Gelman06}. However, in our previous work \citep{Davies17_COST} we found no significant differences in the results from using the more complex prior recommended in that paper.



\section{Posterior Inference}
\label{sec:post_inf}

In order to explore the posterior distribution of the parameters of the eSABRE model we use an MCMC algorithm. Having chosen conjugate priors where possible means we can run a Gibbs sampler for the majority of parameters \citep{Ripley79,GemanGeman84}. These are derived in Section~1 of the online supplementary materials and given here, where by a slight abuse of notation $\bftheta'$ denotes all the other parameters, excluding the ones on the left of the conditioning bar. The only exception is $\bfgam$, which is discussed in Section~\ref{sec:sampling}.
\begin{align}
\boldsymbol\mu_{\textbf{y}}| \bftheta', \bfXg^*, \textbf{Z}, \bfy &\sim \mathcal{N}(\boldsymbol\mu_{\textbf{y}} | \textbf{V}_\textbf{y} (\textbf{M}^\top(\textbf{y} - \textbf{Z}\textbf{b})/\sigma_y^2 + \bfXg^*\bfwg^*/\sigma^2_\varepsilon),\textbf{V}_{\textbf{y}}) \\
\bfwg^* | \bftheta', \bfXg^*, \textbf{Z}, \bfy &\sim \mathcal{N}(\bfwg^* | \textbf{V}_{\bfwg^*} \bfXg^{*\top} \boldsymbol\mu_{\textbf{y}} + \textbf{V}_{\bfwg^*} \bfSigma_{\bfwg^*}^{-1}\textbf{m}_{\bfgam}, \sigma_\varepsilon^2 \textbf{V}_{\bfwg^*}) \\
\textbf{b} | \bftheta', \bfXg^*, \textbf{Z}, \bfy &\sim \mathcal{N}(\textbf{b} | \tfrac{1}{\sigma_y^2}\textbf{V}_{\textbf{b}} \textbf{Z}^\top (\bfy - \textbf{M}\boldsymbol\mu_{\textbf{y}}), \textbf{V}_{\textbf{b}}) \\
\mu_w | \bftheta', \bfXg^*, \textbf{Z}, \bfy &\sim \mathcal{N}(\mu_w | V_{\mu_w} (\textbf{1} \bfwg / \sigma_w^2 + \mu_0 / \sigma_0^2),\sigma_\varepsilon^2 V_{\mu_w}) \\
\sigma_y^2 | \bftheta', \bfXg^*, \textbf{Z}, \bfy &\sim \mathcal{IG}(\sigma_y^2 | ~ ||\bfy||/2 + \alpha_y, \tfrac{1}{2} (\textbf{y} - \textbf{M} \boldsymbol\mu_{\textbf{y}} - \textbf{Z} \textbf{b})^\top (\textbf{y} - \textbf{M} \boldsymbol\mu_{\textbf{y}} - \textbf{Z} \textbf{b})) \\
\sigma_w^2 | \bftheta', \bfXg^*, \textbf{Z}, \bfy &\sim \mathcal{IG}(\sigma_w^2 | ~ ||\bfwg||/2 + \alpha_w, \tfrac{1}{2 \sigma_\varepsilon^2} (\bfwg - \textbf{I} \mu_w)^\top (\bfwg - \textbf{I} \mu_w)) \\
\sigma_{b,g}^2 | \bftheta', \bfXg^*, \textbf{Z}, \bfy &\sim \mathcal{IG}(\sigma_{b,g}^2 | ~ ||\textbf{b}_g||/2 + \alpha_{b,g}, \beta_{b,g} + \tfrac{1}{2} \textbf{b}_g^\top \textbf{b}_g) \\
\sigma_{\varepsilon}^2 | \bftheta', \bfXg^*, \textbf{Z}, \bfy &\sim \mathcal{IG}(\sigma_{\varepsilon}^2 | (||\boldsymbol\mu_{\textbf{y}}|| + ||\bfwg^*|| + 1)/2 + \alpha_\varepsilon, \beta_\varepsilon + \tfrac{1}{2}R_{\sigma_\varepsilon^2}) \\
\pi | \bftheta', \bfXg^*, \textbf{Z}, \bfy &\sim \beta(\pi|~ \alpha_\pi + ||\bfgam||, \beta_\pi + J-||\bfgam|). \label{eq:pi}
\end{align}
where we sample $\sigma_{b,g}^2$ for each $g$. We also define $\textbf{V}_{\textbf{y}} = (1/\sigma_\varepsilon^2 \textbf{I} + \textbf{M}^\top \textbf{M}/\sigma_y^2)^{-1}$, $\textbf{V}_{\bfwg^*} = (\bfXg^{*\top} \bfXg^* + \bfSigma_{\bfwg^*}^{-1})^{-1}$, $\textbf{V}_{\textbf{b}} = (\tfrac{1}{\sigma_y^2 }\textbf{Z}^\top \textbf{Z} + \bfSigma_{\textbf{b}}^{-1})^{-1}$, $V_{\mu_w} = (1/\sigma_0^2 + ||\bfwg||/\sigma_w^2)^{-1}$ and $R_{\sigma_\varepsilon^2} = (\boldsymbol\mu_{\textbf{y}} - \bfXg^* \bfwg^*)^\top (\boldsymbol\mu_{\textbf{y}} - \bfXg^* \bfwg^*) + (\bfwg^* - \textbf{m}_{\bfgam})^\top \bfSigma_{\bfwg^*}^{-1} (\bfwg^* - \textbf{m}_{\bfgam}) + (\mu_{\textbf{w}} - \mu_{0})^\top (\mu_{\textbf{w}} - \mu_{0}) / \sigma_0^2$ for notational simplicity.

Collapsing can lead to improved mixing and convergence, e.g. \cite{AndrieuDoucet99}. We take advantage of the induced conjugacy to sample the parameters $\bfgam$, $\bfwg^*$, $\mu_{w}$, $\sigma_\varepsilon^2$ and $\pi$ as a series of collapsed distributions rather than through Gibbs sampling:
\begin{align}
\label{eq:cond1}
p(\bfgam, \bfwg^*,\mu_w,\sigma_\varepsilon^2,\pi) &= p(\bfgam) p(\pi | \bfgam) p(\sigma_\varepsilon^2 |\pi, \bfgam) p(\mu_w | \sigma_\varepsilon^2, \pi, \bfgam) p(\bfwg^* | \mu_w, \sigma_\varepsilon^2, \pi, \bfgam) \\
\label{eq:cond2}
&= p(\bfgam) p(\pi | \bfgam) p(\sigma_\varepsilon^2 |\bfgam) p(\mu_w | \sigma_\varepsilon^2, \bfgam) p(\bfwg^* | \mu_w, \sigma_\varepsilon^2,\bfgam)
\end{align}
where the conditionality on $\bftheta'$, $\textbf{X}$, $\textbf{Z}$ and $\bfy$ has been dropped in the notation and the simplification from \eqref{eq:cond1} to \eqref{eq:cond2} follows from the conditional independence relations shown in Figure~\ref{fig:extDAG}, exploiting the fact that $\pi$ is d-separated from the remaining parameters in the argument via $\bfgam$. These distributions can be found by collapsing over parameters as derived in Section~1.2 of the online supplementary materials.

\subsection{Sampling the Latent Indicators}
\label{sec:sampling}

Sampling $\bfgam$ is computationally expensive, as a result of it not naturally taking a distribution of standard form. However a conditional distribution can still be obtained and \cite{Davies14,Davies17_COST} used collapsing methods following \cite{SabattiJames05} to achieve faster mixing and convergence as follows:
\begin{align}
p(\bfgam | &\bftheta', \bfXg^*, \textbf{Z}, \textbf{y}) \propto \int p(\bfgam,\pi,\sigma_\varepsilon^2, \bfwg^*,\mu_w| \bftheta', \bfXg^*,\textbf{Z}, \textbf{y}) d\mu_w d\bfwg^* d\pi d\sigma_\varepsilon^2  
\label{eq:gam_post_SABRE}
\end{align}
using the likelihood for the conjugate SABRE model given in \eqref{eq:SABRE_like} and the same priors that are used for the eSABRE model. The closed form solution of this integral can be found in Section~1.1 of the online supplementary materials.

However with the likelihood for the conjugate SABRE model given in \eqref{eq:SABRE_like} the computational cost of computing \eqref{eq:gam_post_SABRE} becomes dependant on inverting a $||\textbf{y}||\times||\textbf{y}||$ matrix. This is a result of integrating over $\bfwg^*$ to give a multivariate Gaussian distribution of dimension $||\textbf{y}||$, which requires a $||\textbf{y}||\times||\textbf{y}||$ matrix inversion in order to evaluate its density and therefore the final density given in \eqref{eq:gam_post_SABRE}. For the size of the datasets used in \cite{Davies14,Davies17_COST} this is not problematic, $||\textbf{y}||=246$ for example. However with the H1N1 dataset, where $||\textbf{y}||=15,693$, calculating any distribution where a $||\textbf{y}||\times||\textbf{y}||$ matrix inversion is repeatedly required becomes practically non-viable.

It is at this point that the structure of the two main probability distributions of the eSABRE model, \eqref{eq:like_lat1} and \eqref{eq:like_lat2}, show the huge computational advantage of the eSABRE model over the conjugate SABRE model proposed in \cite{Davies17_COST}; see Table~\ref{tab:M2_AUROC} for an example of the improved computational efficiency. As in the conjugate SABRE model we use collapsing methods and collapse over $\mu_w$, $\bfwg^*$, $\pi$ and $\sigma_\varepsilon^2$. However while the integration over $\bfwg^*$ in the conjugate SABRE model gives a multivariate Gaussian distribution of size $||\textbf{y}||$ and a computational dependence on the inversion of a $||\textbf{y}||\times||\textbf{y}||$ matrix, where $||\textbf{y}||=15,693$ for the H1N1 dataset. For the eSABRE model we instead get a multivariate Gaussian distribution of dimension $||\boldsymbol\mu_{\textbf{y}}||$ after integrating over $\bfwg^*$, where the evaluation of the density is dependent on a $||\boldsymbol\mu_{\textbf{y}}|| \times ||\boldsymbol\mu_{\textbf{y}}||$ matrix inversion:
\begin{align}
p(\bfgam | &\bftheta', \bfXg^*, \boldsymbol\mu_{\textbf{y}}) \propto \int p(\bfgam,\pi,\sigma_\varepsilon^2, \bfwg^*,\mu_w| \bftheta', \bfXg^*,\boldsymbol\mu_{\textbf{y}}) d\mu_w d\bfwg^* d\pi d\sigma_\varepsilon^2.  
\label{eq:gam_post_eSABRE}
\end{align}
This dependence on $||\boldsymbol\mu_{\textbf{y}}||$ rather than $||\textbf{y}||$ is where the main computational cost reduction occurs, as in the H1N1 dataset $||\boldsymbol\mu_{\textbf{y}}||=570$ is much smaller than $||\textbf{y}||=15,693$ making the computational cost of computing \eqref{eq:gam_post_eSABRE} far less than \eqref{eq:gam_post_SABRE}. It is this reduction in computational costs that makes the eSABRE model feasible for the H1N1 dataset, where the computational cost of the SABRE models is prohibitive.

Multiple methods have been proposed for sampling the latent variables, $\bfgam$. \cite{Davies14} looked at two of these in particular; the component-wise Gibbs sampling approach and a Metropolis-Hastings step \citep{Metropolis53,Hastings70}. In those studies it was found that block Metropolis-Hastings sampling was the method that offered the quickest convergence of the model based on CPU time and we have therefore used this method here.

Block Metropolis-Hastings sampling improves mixing and convergence through proposing sets, $S$, of latent indicator variables, $\bfgam_S$, simultaneously, where $\bfgam_S$ denotes a column vector of all the $\gamma_j$s where $j \in S$ and $\bfgam_{-S}$ its compliment. The proposals are then accepted with the following acceptance rate based on the current state, $c$ of all the other $\gamma$s:
\begin{align}
	\label{eq:MH}
	\alpha(\bfgam_S^*,\bfgam_S^{c} | \bftheta', \bfXg^*, \textbf{Z}, \bfy, \bfgam_{-S}^c) := \min\left\{ \frac{q(\bfgam_S^{c} | \bfgam_S^*,\pi) p(\bfgam_S=\bfgam_S^*,\bfgam_{-S}^c|  \bftheta', \bfXg^*, \textbf{Z}, \bfy)}{q(\bfgam_S^* | \bfgam_S^{c},\pi) p(\bfgam_S=\bfgam_S^c,\bfgam_{-S}^c| \bftheta', \bfXg^*, \textbf{Z}, \bfy)},1 \right\}
\end{align}
where $q(.)$ is a proposal density, which we set to be:	$q(\bfgam_S^* | \bfgam_S^{c},\pi) = \prod_{j \in S} \Bern(\gamma_j^* | \pi)$.For the SABRE model \eqref{eq:gam_post_SABRE} is used for computing $p(.)$ in \eqref{eq:MH}, while \eqref{eq:gam_post_eSABRE} is used for the eSABRE model. Proposed moves for independent sets of randomly ordered inclusion parameters, $\bfgam_S^*$, are then accepted if $\alpha(\bfgam_S^*,\bfgam_S^{c} | \bftheta', \bfXg^*, \textbf{Z}, \bfy, \bfgam_{-S}^c)$ is greater than a random variable $u \sim \mathcal{U}[0,1]$, until updates have been proposed for all the latent indicator variables.

\section{Selection of Random Effect Components}
\label{sec:M2_SelectingRE}

There are various methods that can be used to select the random effects that should be used within a model. Previously \cite{Davies16_LNBI} compared 10-fold Bayesian CV and WAIC \citep{Watanabe10}, and found that in terms of model selection WAIC achieved a similar performance at a lower computational cost to 10-fold Bayesian CV. Here we look at Bayesian integrated CV (iCV), e.g. \cite{VehtariOjanen12}, and several variations of WAIC that can be applied to latent variable models. 

An alternative approach to those suggested above would be to use spike and slab priors to select the random effects. While this would only require one model to be fitted, doing so will come at a large computational cost. This is a result of poor mixing associated with proposing MCMC moves which change entire groups of random effect coefficients simultaneously. Using intra-model approaches for a small number of models in parallel is far more computationally viable and has therefore been used here.

\subsection{Integrated Cross Validation}
\label{sec:M2_iCV}

Bayesian CV methods are reliable, if computationally expensive, techniques for measuring the out-of-sample performance of different models. Bayesian iCV, e.g. \cite{VehtariOjanen12}, is a special version of CV which works well in latent variable models. Bayesian iCV integrates over the latent variables, in this case $\boldsymbol\mu_{\textbf{y}}$, to give the following utility function for k-fold Bayesian iCV:
\begin{align}
\label{eq:iCV}
p_{iCV} = \frac{1}{K} \sum_{k=1}^{K} \log \frac{1}{I} \sum_{\iota=1}^{I}  N(\textbf{y}_{k} | \textbf{M}_k \textbf{X}_{\bfgam,k}^* \textbf{w}_{\bfgam}^{*,\iota} +\textbf{Z}_k \textbf{b}^\iota, \sigma_y^{\iota,2} \textbf{I}_k + \sigma_\varepsilon^{\iota,2} \textbf{M}_k \textbf{M}_k^\top)
\end{align}
where the distribution comes from integrating over $\boldsymbol\mu_{\textbf{y}}$ in the distribution given by the product of \eqref{eq:like_lat1} and \eqref{eq:like_lat2}. The parameter samples, $\textbf{w}_{\bfgam}^{*,\iota}$, $\textbf{b}^\iota$, $\sigma_y^{\iota,2}$ and $\sigma_\varepsilon^{\iota,2} $, are taken from the eSABRE model applied to $\textbf{y}_{-k}$, $\textbf{X}_{-k}$, $\textbf{Z}_{-k}$ and $\textbf{M}_{-k}$.

\subsection{Block Integrated WAIC}
\label{sec:M2_biWAIC}

WAIC, as proposed in \cite{Watanabe10}, is a natural method for selecting the correct model when the underlying model is singular, i.e models with a non-identifiable parameterisation, such as the SABRE and eSABRE models. WAIC has been proven to be asymptotically equivalent to Bayesian leave-one-out CV (LOO-CV) in \cite{Watanabe10} and is computed as follows from posterior samples $\boldsymbol\theta^\iota$ for $\iota \in \{1,\dots,I\}$:
\begin{align}
	\label{eq:M2_WAIC}
	p_{WAIC} &= -2 \sum_{i=1}^N \bigg( \log \left( \frac{1}{I} \sum_{\iota = 1}^{I} p(y_i | \boldsymbol\theta^\iota,\textbf{X}_{\bfgam,i},\textbf{Z}_i) \right) - \Var \left(\log (p(y_i | \boldsymbol\theta^\iota,\textbf{X}_{\bfgam,i},\textbf{Z}_i) ) \right) \bigg). 
\end{align}
where $\Var$ is the sample variance with respect to $\boldsymbol\theta^\iota$. WAIC can be used for a wide variety of problems, however it is only justifiable for problems where the observed data are independently distributed with a population distribution, e.g. the SABRE model where the joint likelihood is given by \eqref{eq:SABRE_like}. The inclusion of latent variables in the eSABRE model means that the observed data are not modelled with independent distributions and it is therefore inaccurate to use WAIC with the eSABRE model.

To make WAIC more applicable to latent variable models such as the eSABRE model, \cite{Li15} introduced two alternative versions of WAIC; non-integrated WAIC (nWAIC) and integrated WAIC (iWAIC). nWAIC applies WAIC to the predictive density of the observed variables, $\bfy = (y_1,\ldots,y_N)$, conditional on the model parameters, $\boldsymbol\theta$, and the potentially correlated latent variables, $\boldsymbol\psi = (\psi_1,\ldots,\psi_N)$:
\begin{align}
	\label{eq:M2_nWAIC}
	p_{nWAIC} &= -2 \sum_{i=1}^N \bigg( \log \left( \frac{1}{I} \sum_{\iota = 1}^{I} p(y_i | \boldsymbol\theta^\iota,\psi_i^\iota,\textbf{Z}_i) \right) - \Var \left( \log ( p(y_i | \boldsymbol\theta^\iota,\psi_i^\iota,\textbf{Z}_i) ) \right) \bigg)
\end{align}
where $\boldsymbol\theta^\iota$ and $\psi_i^\iota$ are sampled from the posterior distribution via MCMC and $\Var$ is the sample variance. In the proposed eSABRE model, taking just the likelihood for $y_i$ from \eqref{eq:like_lat1} would be the distribution corresponding to $p(y_i | \boldsymbol\theta^\iota,\psi_i^\iota,\textbf{Z}_i)$ and would not satisfy the independence assumptions of WAIC based methods as each $y_i$ is dependent on a latent variable, $\psi_i$, which is shared by other observations.

nWAIC also does not fully account for the mismatch in the model fit of the latent variables, i.e. how well the latent variables are predicted by the fixed effects. \cite{Li15} therefore proposed iWAIC:
\begin{align}
	\label{eq:M2_iWAIC}
	p_{iWAIC} &= -2 \sum_{i=1}^N \bigg( \log \left( \frac{1}{I} \sum_{\iota = 1}^{I} p(y_i | \boldsymbol\theta^\iota,\textbf{X}_{\bfgam,i},\textbf{Z}_i,\boldsymbol\psi_{\textbf{-i}}^\iota) \right) - \Var \left( \log (p(y_i | \boldsymbol\theta^\iota,\textbf{X}_{\bfgam,i},\textbf{Z}_i,\boldsymbol\psi_{\textbf{-i}}^\iota)) \right) \bigg)
\end{align}
where $\Var$ is the sample variance and the distribution used is given by $p(y_i | \boldsymbol\theta^\iota,\textbf{X}_{\bfgam,i},\textbf{Z},\boldsymbol\psi_{\textbf{-i}}^\iota)$ $= \int p(y_i | \boldsymbol\theta^\iota,\boldsymbol\psi_{\textbf{-i}}^\iota,\psi_i,\textbf{Z})p(\psi_i | \boldsymbol\theta^\iota,\bfXg)d\psi_i$, the analytical integration of the latent variables from the product of the likelihood and the distribution of the latent variables.

The proposed version of iWAIC does not however work with the eSABRE model. This is a result of each observation, $y_i$, not having its own corresponding latent variable, $\psi_i$. Instead any two observations, $y_1$ and $y_2$, from the same pair of reference and test viruses, $p$, will have the same latent variable, i.e. $\psi_1=\psi_2=\mu_{\textbf{y},p}$. Under this model, i.e. where $\rho(\psi_1,\psi_2)=1$, it is mathematically intractable to integrate over $\psi_1 = \mu_{\textbf{y},p}$ without integrating over $\psi_2 = \mu_{\textbf{y},p}$, something which is required in order to calculate $p(y_i | \boldsymbol\theta^\iota,\textbf{X}_{\bfgam,i},\textbf{Z}_i,\boldsymbol\psi_{\textbf{-i}})$ as needed for \eqref{eq:M2_iWAIC}. We must therefore either use nWAIC given by \eqref{eq:M2_nWAIC} or find an alternative.

In this current work we propose biWAIC, a new modification of WAIC for latent variable models with latent variables that are either completely correlated or have no correlation, like in the eSABRE model. While WAIC, nWAIC and iWAIC rely on using independent distributions for each $y_i$, biWAIC instead uses a distribution for independent groups of observations $\textbf{y}_p$ with the same associated latent variable, such that $\textbf{y}_p$ is the group containing all $y_i$ whose virus pair, $p_i$, is the same as the groups virus pair, $p$. Given this notation we can then compute biWAIC as follows:
\begin{align}
	\label{eq:M2_biWAIC}
	p_{biWAIC} &= -2 \sum_{p=1}^P \bigg( \log \left( \frac{1}{I} \sum_{\iota = 1}^{I} p(\textbf{y}_p | \boldsymbol\theta^\iota,\textbf{X}_{\bfgam,p},\textbf{Z}_p) \right) - \Var \left( \log ( p(\textbf{y}_p | \boldsymbol\theta^\iota,\textbf{X}_{\bfgam,p},\textbf{Z}_p) ) \right) \bigg)
\end{align}
where $\Var$ is the sample variance and the distribution used is the analytic integration over the corresponding latent variable, $\mu_{y,p}$, of the product of the likelihood and the distribution of the latent variables taken from \eqref{eq:like_lat1} and \eqref{eq:like_lat2}: $p(\textbf{y}_p | \boldsymbol\theta^\iota,\textbf{X}_{\bfgam,p},\textbf{Z}) = \int p(\textbf{y}_p | \boldsymbol\theta^\iota,\mu_{y,p},\textbf{Z})$ $p(\mu_{y,p} | \boldsymbol\theta^\iota,\textbf{X}_{\bfgam,p}) d\mu_{y,p}$.

As well as being applicable to the eSABRE model and particular specifications of latent variable models, biWAIC can also be shown to have some useful asymptotic properties. Previously \cite{Watanabe10} has shown that WAIC is asymptotically equivalent to Bayesian leave-one-out CV (LOO-CV), based on the fact that Bayesian LOO-CV loss is asymptotically equivalent to WAIC as a random variable. While biWAIC is not asymtotically equivalent to Bayesian LOO-CV, based on the same proof of \cite{Watanabe10} we can determine that it is asymptotically equivalent to a different form of Bayesian CV. From looking at \eqref{eq:iCV} and \eqref{eq:M2_biWAIC}, along with the two distributions from which those equations are derived, \eqref{eq:like_lat1} and \eqref{eq:like_lat2}, we can see that if iCV is evaluated on the same groups as biWAIC, then it is asymptotically equivalent to biWAIC as a random variable. biWAIC is therefore asymptotically equivalent to Bayesian leave-one-group-out CV (LOGO-CV), where observations are divided into $P$ independent groups based on the number of different virus pairs (groups), as opposed to $n$ groups of single observations for Bayesian LOO-CV.


\section{Simulated Datasets}
\label{sec:Sim_Data}

In this section we describe the simulated datasets that are used to test the effectiveness of the eSABRE model proposed here and compare it with the conjugate SABRE model described in \cite{Davies16_LNBI}

\subsection{Non-FMDV Simulated Data}
\label{sec:M2_SimData_NON_FMDV}

To initially test the eSABRE and conjugate SABRE models we generated 3 datasets with a reasonably small number of variables. These 3 datasets (Simulated Dataset 1 (SD1), SD2 and SD3) are based on the same structure as the influenza datasets with a varied number of random effect factors. In each of the datasets 2000 observations were simulated from 55 pairs of viruses. The 55 pairs of viruses comes from having 10 viruses tested against each other (${10 \choose 2} = 45$) plus the viruses tested against themselves (10), with each of these pairs then given 50 possible fixed effects and 4 possible random effect components (including the reference and test viruses). The random effects groups were included with probability 0.5 and given zero coefficients otherwise, while the relevant coefficients were generated from a zero mean Gaussian distribution with each component having a fixed variance drawn from $U(0.2,0.5)$. Fixed effects, $w_j$, were given non-zero values generated from a uniform distribution, $U(-0.4,-0.2)$, with inclusion probability $\pi \sim U(0.2,0.4)$. $\sigma_y^2$ and $\sigma_\varepsilon^2$ were both set to be 0.033, 0.1 and 0.3 respectively for the three simulated datasets.

\subsection{FMDV Simulated Data}
\label{sec:M2_SimData_FMDV}

To make the simulation studies more realistic we wanted to make simulated datasets based on the influenza A(H1N1) dataset described in Section~\ref{sec:H1N1}. However, while this does not cause any problems for the proposed eSABRE model, using the conjugate SABRE model to analyse datasets of this size is computationally prohibitive. Therefore instead we have created 20 simulated datasets based on the extended South African Territories (SAT) type 1 FMDV dataset used in \cite{Reeve16} and \cite{Davies17_COST}. These datasets were created to be the same size as the FMDV datasets using the maximum a-posteriori parameter estimates of the eSABRE method applied to the SAT1 FMDV dataset. However in order to highlight the differences in performance of the two models under different circumstances, we varied the error of the underlying model, $\sigma_\varepsilon^2 \in \{0.02,0.2,0.5\}$, and changed the mean of the regression parameters, $\mu_w \in \{-0.1,-0.3,-0.5\}$. Following \cite{Reeve16} we used 3 random effect components; the test virus, the date of the experiment and the antiserum (reference virus).

\subsection{Simulated Data for Model Selection}
\label{sec:M2_SimData_ModelSelection}

Finally, to compare nWAIC, biWAIC and 10-fold Bayesian iCV, we have generated 9 sets of 20 datasets with up to 4 random effects; the test virus, the reference virus and two generic random effect factors. The datasets were generated with 50 possible fixed effects and up to 4 random effect factors included with probability 0.5. Of the 9 sets of datasets, 3 contain 10 virus strains, where each virus strain has been used as a both a reference and test virus, meaning there are 55 pairs of reference and test viruses; see Section~\ref{sec:M2_SimData_NON_FMDV}. Following the same set up, 3 of the sets of datasets include 30 virus strains (465 pairs) and the other 3 have 45 virus strains (1035 pairs). Within each of these sets of 3 datasets, the model error, $\sigma_\varepsilon^2$, was varied to be either 0.1, 0.3 or 0.5.

\section{Computational Inference}
\label{sec:M2_Simulations}

To test the convergence of the parameter samples for both the simulated and real datasets we generated 4 chains of parameters samples for each model and then computed the PSRF \citep{GelmanRubin92} from the within-chain and between-chain variances. We took the threshold of convergence to be a PSRF $\leq1.1$ and terminated the burn-in phase when this was satisfied for 95\% of the variables. The fixed hyperparameters, shown as grey nodes in Figure~\ref{fig:extDAG}, were set the same for both the eSABRE and conjugate SABRE methods such that $\boldsymbol\alpha_{\textbf{b}}=\boldsymbol\beta_{\textbf{b}}=(0.001,\ldots,0.001)$, $\alpha_w=\beta_w=\alpha_y=\beta_y=\alpha_\varepsilon=\beta_\varepsilon=0.001$, $\mu_0=0$, $\sigma_0^2=100$, $w_0=max(y)$, $\alpha_\pi=1$ and $\beta_\pi=4$ following \cite{Davies17_COST}.

\begin{table}[t]
				\centering
				\caption[Table of AUROC values and CPU time for the eSABRE and the conjugate SABRE models applied to the non-FMDV based simulated datasets.]{\textbf{Table of AUROC values and CPU time for the eSABRE and the conjugate SABRE models applied to the non-FMDV based simulated datasets.} The table gives the AUROC values and CPU times per 1,000 iterations (seconds) for the eSABRE and conjugate SABRE models, where the results for the conjugate SABRE model are given in brackets. The result come from when the methods were applied to the non-FMDV simulated datasets (SD1, SD2 and SD3) described in Section~\ref{sec:M2_SimData_NON_FMDV} with varied numbers of observations.}
				\begin{tabular}{c | c  c  c | c  c  c }
			\multirow{2}{*}{\textbf{Obs.}}	& \multicolumn{3}{|c}{\textbf{AUROC Values}} & \multicolumn{3}{|c}{\textbf{CPU Time Per 1,000 Iterations}} \\ \cline{2-7}
										  & \textbf{SD1} & \textbf{SD2} & \textbf{SD3} & \textbf{SD1} & \textbf{SD2} & \textbf{SD3} \\ \hline
										\textbf{500} & 0.98 (0.90) & 0.90 (0.77) & 0.82 (0.64) & 25 (497) & 25 (867) & 47 (444) \\
										\textbf{1000} & 0.98 (0.83) & 0.91 (0.70) & 0.82 (0.59) & 29 (6,931) & 26 (5,623) & 36 (5,546) \\
										\textbf{2000} & 0.98 (0.75) & 0.92 (0.61) & 0.83 (0.58) & 32 (35,231) & 25 (32,243) & 43 (20,904)
 				\end{tabular}
        \label{tab:M2_AUROC}	
\end{table}

\section{Results for the Simulation Studies}
\label{sec:M2_SimulationStudies}

Table~\ref{tab:M2_AUROC} gives the Area Under the Reciever Operating Characteric curve (AUROC) values for the eSABRE model proposed here and the conjugate SABRE model described in \cite{Davies17_COST} applied to the non-FMDV simulated datasets from Section~\ref{sec:M2_SimData_NON_FMDV}; SD1, SD2, SD3. For each combination of dataset and number of observations, the eSABRE model offers a clear improvement in terms of global variable selection performance over the SABRE model. This improvement is a result of the latent variable structure of the eSABRE model which better reflects the data generation process, where the difference in the models can be seen by comparing the Probabilistic Graphical Models (PGMs) in Figure~\ref{fig:extDAG} here and Figure~1 in \cite{Davies17_COST}. Table~\ref{tab:M2_AUROC} also shows how this improvement is more significant as the effect of the latent variable structure is increased, i.e. as $\sigma_\varepsilon^2$ is increased. As the error variances get larger, e.g. SD2 and SD3, the eSABRE model offers a much clearer improvement over the conjugate SABRE model than it did it in SD1 dataset where the error variances are smaller. This is a result of the conjugate SABRE and eSABRE models becoming more similar as $\sigma_\varepsilon^2 \rightarrow 0$. Given the large variance in HI assay measurements for any given pair of reference and test viruses in the H1N1 dataset, this improvement is vital.

\begin{figure}[t]
				\vspace{-1cm}
				\centerline{\includegraphics[width=0.8\textwidth]{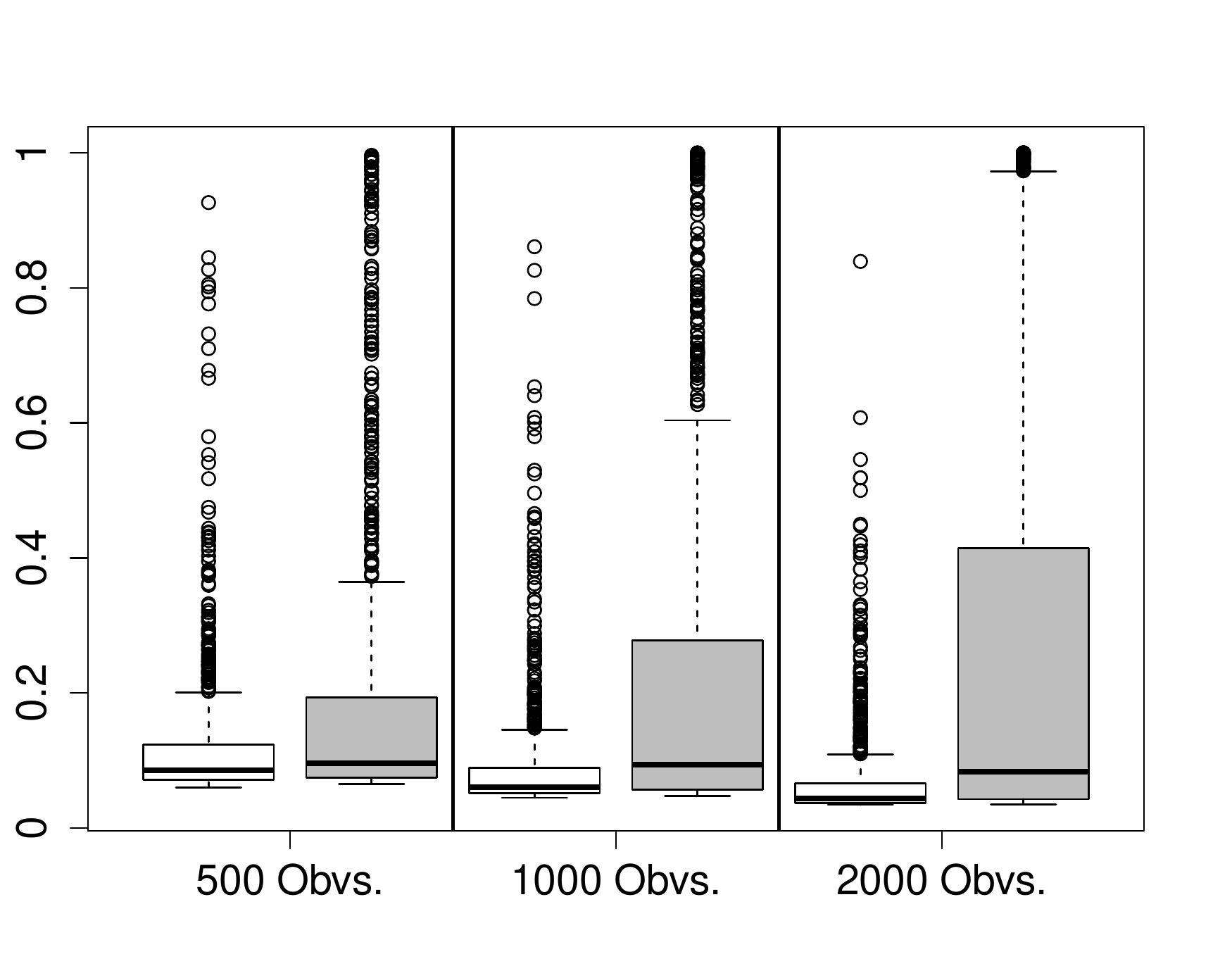}}
				\vspace{-0.7cm}
				\caption[Box plots showing the effect of non-iid Gaussian noise on a model assuming iid Gaussian noise.]{\textbf{Box plots showing the effect of non-iid Gaussian noise on a model assuming iid Gaussian noise.} The box plots show the probability of an irrelevant variable being included in a model for data with iid Gaussian noise (white) against the probabilities for a model with a noise structure based on the H1N1 dataset (grey). The results show the probability of the irrelevant variable being included in the model decrease as the number of observations increases for the data with iid Gaussian noise. Conversely it shows an increase in the probability of its inclusion as the number of observations increases when there is a noise structure based on the H1N1 dataset.}
				\label{fig:M2_Boxplots}
\end{figure}

Another notable result from Table~\ref{tab:M2_AUROC} is the reduction in performance in terms of AUROC values of the conjugate SABRE model as the number of observations increases. This is an unexpected result as we would expect more data to provide more information to the model, resulting in a better selection of variables in the models and higher AUROC values. The reason for this unexpected result is a consequence of the mismatch between the data generation process where variance in the observations comes in two forms, $\sigma^2_\varepsilon$ and $\sigma^2_y$, and the model which only directly accounts for the variance in $\bfy$ given by $\sigma^2_y$.

To demonstrate that the unexpected reduction in performance of the conjugate SABRE model is a result of the mismatch between the data and the model we have completed a small simulation study with linear models. We have generated groups of datasets with 500, 1,000 and 2,000 observations generated from a linear model with each group containing 2000 datasets. For each of these groups, half the datasets have observations generated with iid noise, e.g. $\textbf{y} \sim \mathcal{N}(\textbf{y}|\textbf{X}\textbf{w},\sigma_y^2 \textbf{I})$. The other half of the datasets were given correlated errors based on integrating over a set of random effects, e.g. $\textbf{y} \sim \mathcal{N}(\textbf{y}|\textbf{X}\textbf{w},\sigma_y^2 \textbf{I} + \sigma_\varepsilon^2 \textbf{M} \textbf{M}^\top)$. This is equivalent to integrating over the latent variables, but allows us to use the same $\textbf{X}$ and $\textbf{w}$ for a fair comparison. Additionally each of the datasets contains two variables, one relevant, $\textbf{x}_{r}$, and one irrelevant, $\textbf{x}_{ir}$. We have then calculated the marginal likelihood for each of the four possible models\footnote{ marginal likelihood with (1) no variables included: $p(\textbf{y} | .)$, (2) irrelevant variable included only: $p(\textbf{y} | \textbf{x}_{ir})$, (3) relevant variable included only: $p(\textbf{y} | \textbf{x}_{r})$ and (4) both variables included: $p(\textbf{y} | \textbf{x}_{ir}, \textbf{x}_{r})$.} under the assumption of iid noise, as the conjugate SABRE model assumes (incorrectly) with the H1N1 data, where we have fixed $\sigma_w^2$ and marginalised out $\sigma_y^2$ and $\textbf{w}$. We can then use these marginal likelihoods to calculate the probability of the irrelevant variable being included in the final model, $\mathcal{M}$, via Bayes theorem as follows:
\begin{align}
		\label{eq:M2_Irrelevant}
		\mathbb{P}(\textbf{x}_{ir} \in \mathcal{M}) = \frac{p(\textbf{y} | \textbf{x}_{ir}) + p(\textbf{y} | \textbf{x}_{ir}, \textbf{x}_{r})}{p(\textbf{y} | .) + p(\textbf{y} | \textbf{x}_{ir}) + p(\textbf{y} | \textbf{x}_{r}) + p(\textbf{y} | \textbf{x}_{ir}, \textbf{x}_{r})}.
\end{align}

Figure~\ref{fig:M2_Boxplots} gives box plots of the probability of the irrelevant variable, $\textbf{x}_{ir}$, being included in the final model for each of the datasets from our small simulation study. The box plots show the effect on the probabilities caused by the different types of noise and varied amounts of observations. Figure~\ref{fig:M2_Boxplots} shows that as the number of observations increases the chance of the irrelevant variable being included decreases for the iid noise, as would be expected. However for the non-iid noise based on the FMDV and Influenza datasets, the results show an increase in the probability of the irrelevant variable being included as the number of observations increases, indicating that the model mismatch inherent in the SABRE model is what causes the unexpected results in Table~\ref{tab:M2_AUROC}.

Finally, Table~\ref{tab:M2_AUROC} shows the improvement the eSABRE model offers over the conjugate SABRE model in terms of computational efficiency. Table~\ref{tab:M2_AUROC} shows how the conjugate SABRE model becomes far more computationally expensive as the number of observations increases, while the required CPU time hardly changes for the eSABRE model if the number of pairs of reference and test viruses remains the same. This improvement in terms of computational efficiency explains why it is viable to use the eSABRE model on the H1N1 dataset for example, where $||y||=15,693$ and $P=570$, but not the conjugate SABRE model described in \cite{Davies17_COST}.

\begin{table}[t]
				\centering
				\caption[Table of AUROC values for the eSABRE and the conjugate SABRE models when applied to the FMDV based simulated datasets.]{\textbf{Table of AUROC values for the eSABRE and the conjugate SABRE models when applied to the FMDV based simulated datasets.} The table gives AUROC values for the eSABRE and conjugate SABRE models, where the results for the conjugate SABRE model are given in brackets, when applied to the FMDV based simulated datasets described in Section~\ref{sec:M2_SimData_FMDV}.}
				\begin{tabular}{c  c | c  c  c }
						\multicolumn{2}{c|}{ } 						& \multicolumn{3}{c}{$\sigma_\varepsilon^2$} 	\\ \cline{3-5}
						\multicolumn{2}{c|}{ } 						& 0.02    			& 0.2     		& 0.5  				\\ \hline
						\multirow{3}{*}{$\mu_w$} 	&	-0.1	&	0.67 (0.69)		& 0.67 (0.60)	&	0.63 (0.57) \\
																			&	-0.3	&	0.72 (0.71)		&	0.70 (0.61)	& 0.67 (0.58) \\
																			&	-0.5	& 0.75 (0.72)		& 0.74 (0.64)	& 0.73 (0.57)
				\end{tabular}
        \label{tab:M2_AUROC_FMDV}	
\end{table}

Table~\ref{tab:M2_AUROC_FMDV} shows the effectiveness of the eSABRE model on larger more realistic datasets (Section~\ref{sec:M2_SimData_FMDV}) based on the real life FMDV data from \cite{Reeve16}. Like Table~\ref{tab:M2_AUROC}, the results of Table~\ref{tab:M2_AUROC_FMDV} again show the eSABRE model clearly outperforming the conjugate SABRE model across all of the simulated datasets from Section~\ref{sec:M2_SimData_FMDV}. The results show that as the model error in the simulated data increases, the conjugate SABRE model seriously drops off in performance while the eSABRE model remains reasonably consistent. Like with the results of Table~\ref{tab:M2_AUROC}, the difference in performance is again caused by the mismatch between the conjugate SABRE model and the underlying data generation process, which the eSABRE model matches more closely.

\begin{table}[p]
			\caption{\textbf{Table of results looking at the random effects factor selection performance of the methods described in Section~\ref{sec:M2_SelectingRE}.} The table gives results in terms of the successful selection or exclusion of random effects factors when using the methods described in Section~\ref{sec:M2_SelectingRE}, nWAIC, biWAIC and Bayesian 10-fold iCV, on parameter samples from the posterior distribution of the eSABRE model applied to the simulated data from Section~\ref{sec:M2_SimData_ModelSelection}, where $P$ is the number of pairs of reference and test strains. The results given are sensitivity, specificity and F1-scores and are displayed in an alternative manner in Figures~\ref{fig:M2_Barplot} and~\ref{fig:M2_ROC}. F1-scores are described in footnote 2.}
				\centering
				\begin{tabular}{c | c  c | c  c  c }
				 & $P$ & $\sigma_\varepsilon^2$ & nWAIC & biWAIC & Bayesian 10-fold iCV  \\ \hline
				\multirow{9}{*}{\textbf{Sensitivity}} & 55 		& 0.1 & 0.90 & 0.97 & 0.92 \\
																							& 55 		& 0.3 & 0.92 & 0.90 & 0.89 \\
																							& 55 		& 0.5 & 0.78 & 0.71 & 0.93 \\
																							& 465 	& 0.1 & 0.97 & 0.94 & 0.85 \\
																							& 465 	& 0.3 & 0.86 & 0.84 & 0.86 \\
																							& 465 	& 0.5 & 0.95 & 0.90 & 0.86 \\
																							& 1035 	& 0.1 & 0.93 & 0.71 & 0.98\\
																							& 1035 	& 0.3 & 0.91 & 0.79 & 0.87 \\
																							& 1035 	& 0.5 & 0.90 & 0.66 & 0.74 \\ \hline
				\multirow{9}{*}{\textbf{Specificity}} & 55 		& 0.1 & 0.68 & 0.56 & 0.15 \\
																							& 55 		& 0.3 & 0.70 & 0.60 & 0.41 \\
																							& 55 		& 0.5 & 0.59 & 0.54 & 0.26 \\
																							& 465 	& 0.1 & 0.45 & 0.60 & 0.66 \\
																							& 465 	& 0.3 & 0.49 & 0.63 & 0.63 \\
																							& 465 	& 0.5 & 0.37 & 0.56 & 0.53 \\
																							& 1035 	& 0.1 & 0.32 & 0.60 & 0.47 \\
																							& 1035 	& 0.3 & 0.33 & 0.52 & 0.33 \\
																							& 1035 	& 0.5 & 0.39 & 0.55 & 0.29 \\ \hline
				\multirow{9}{*}{\textbf{F-Score}} 		& 55 		& 0.1 & 0.80 & 0.80 & 0.65 \\
																							& 55 		& 0.3 & 0.88 & 0.84 & 0.79 \\
																							& 55 		& 0.5 & 0.72 & 0.66 & 0.70 \\
																							& 465 	& 0.1 & 0.70 & 0.75 & 0.73 \\
																							& 465 	& 0.3 & 0.70 & 0.74 & 0.75 \\
																							& 465 	& 0.5 & 0.73 & 0.76 & 0.72 \\
																							& 1035 	& 0.1 & 0.73 & 0.69 & 0.80 \\
																							& 1035 	& 0.3 & 0.77 & 0.74 & 0.75 \\
																							& 1035 	& 0.5 & 0.60 & 0.54 & 0.60 
				\end{tabular}
				\label{tab:M2_Selection}
\end{table}

\begin{figure}[t]
				\centerline{\includegraphics[width=0.8\textwidth]{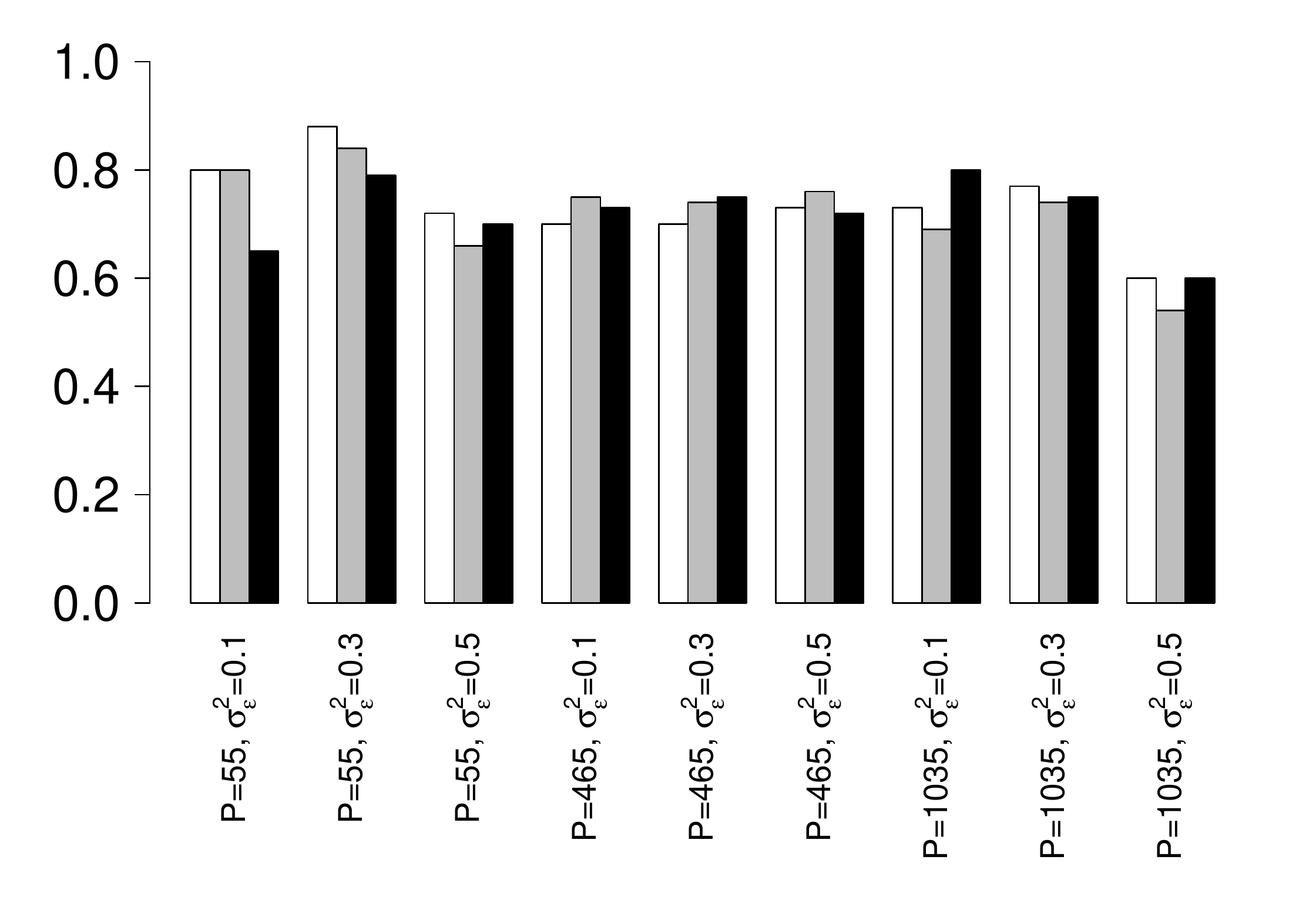}}
				\caption[Bar plot of F1-Scores given in Table~\ref{tab:M2_Selection}.]{\textbf{Bar plot of F1-Scores given in Table~\ref{tab:M2_Selection}.} The bar plot compares the F1-scores for nWAIC (white), biWAIC (grey) and Bayesian 10-fold iCV (black) in terms of correctly selecting random effect components for the dataset described in Section~\ref{sec:M2_SimData_ModelSelection}. The figure takes the results from Table~\ref{tab:M2_Selection}.}
				\label{fig:M2_Barplot}
\end{figure}

\begin{figure}[t]
				\centerline{\includegraphics[width=0.55\textwidth]{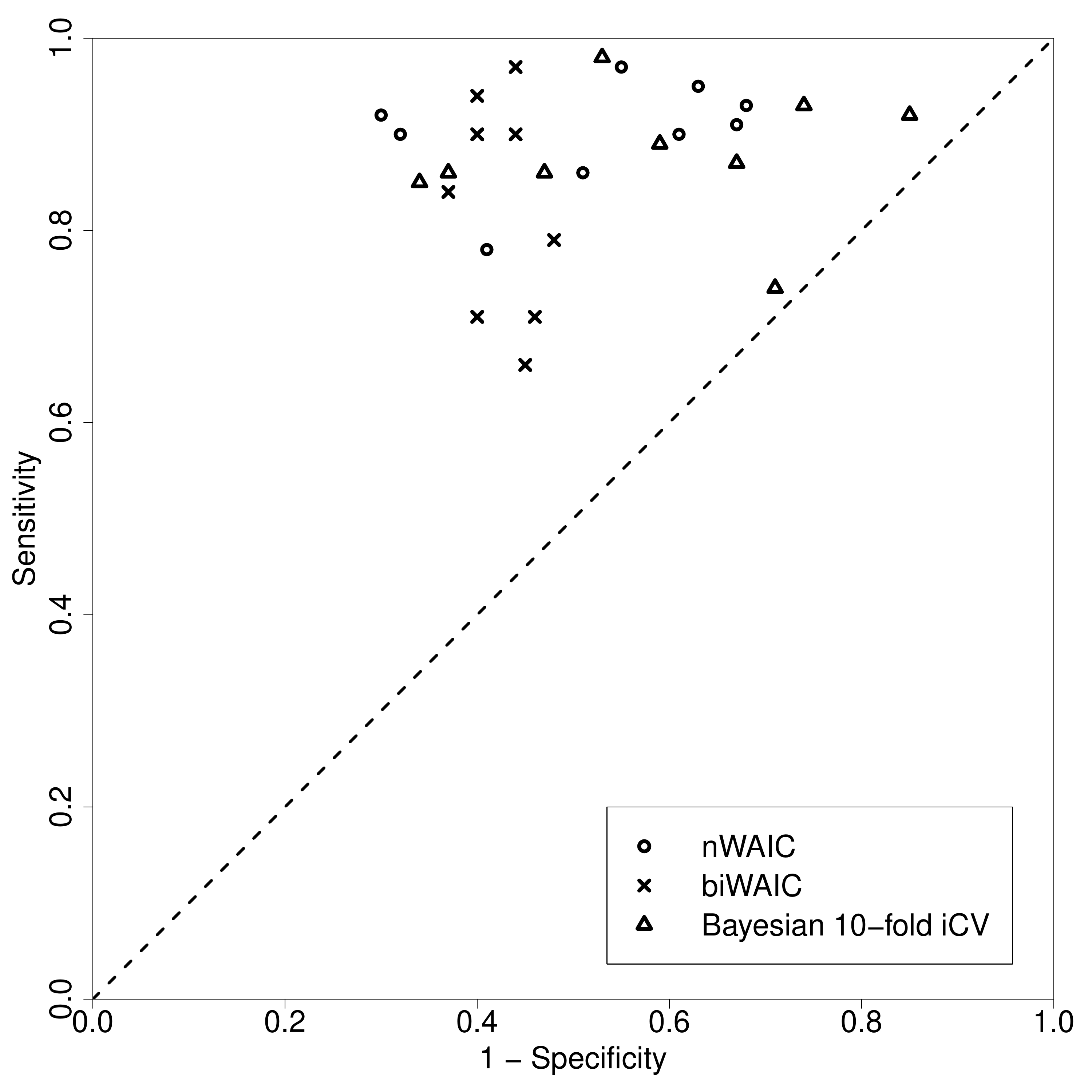}}
				\caption[Plot of sensitivities and complementary specificities for the results given in Table~\ref{tab:M2_Selection}.]{\textbf{Plot of sensitivities and 1 minus specificities for the results given in Table~\ref{tab:M2_Selection}.} The plot compares nWAIC (circles), biWAIC (crosses) and Bayesian 10-fold iCV (triangles) in terms of correctly selecting random effect components for the dataset described in Section~\ref{sec:M2_SimData_ModelSelection}. The figure takes the results from Table~\ref{tab:M2_Selection} and plots the sensitivities against the complementary specificities (i.e i minus specificities), i.e. as single point from a ROC curve.}
				\label{fig:M2_ROC}
\end{figure}

To compare the methods described in Section~\ref{sec:M2_SelectingRE}, nWAIC, biWAIC and Bayesian 10-fold iCV, we have looked at their performance in terms of correctly selecting random effect factors on the datasets from Section~\ref{sec:M2_SimData_ModelSelection}. The results are given in Table~\ref{tab:M2_Selection} and are displayed visually in Figures~\ref{fig:M2_Barplot} and~\ref{fig:M2_ROC}.

The results in Table~\ref{tab:M2_Selection} show that all of the methods, nWAIC, biWAIC and Bayesian 10-fold iCV, perform similarly in terms of overall selection accuracy. The similarity is best demonstrated by looking at the F1-scores, which consider both precision and recall, offering a more general assessment of performance than looking at them separately.\footnote{F1-scores can be calculated as follows: $F1 = 2 \times \tfrac{Precision \times Recall}{Precision + Recall}$} The F1-scores from Table~\ref{tab:M2_Selection} can also be seen in Figure~\ref{fig:M2_Barplot} where the results are shown as bar plots. With the results from Table~\ref{tab:M2_Selection} and Figure~\ref{fig:M2_Barplot} suggesting that the information criteria, nWAIC and biWAIC, give similar selection performance to Bayesian 10-fold iCV, it is reasonable to use one of the former criteria on the Influenza dataset in Section~\ref{sec:H1N1_results}, where Bayesian 10-fold iCV will be computationally onerous.

While suggesting that the methods perform similarly overall in terms of F1-scores, Table~\ref{tab:M2_Selection} also indicates that the methods operate with different sensitivity versus specificity trade offs, meaning that on average some methods include more random effect factors than others. This can be seen by looking at the sensitivities and specificities of nWAIC, biWAIC and Bayesian 10-fold iCV in Table~\ref{tab:M2_Selection} or alternatively by looking at Figure~\ref{fig:M2_ROC}. Figure~\ref{fig:M2_ROC} plots the sensitivities achieved by the different methods on each set of datasets against the complementary specificity (i.e. 1 minus specificity) and shows that the biWAIC method operates at a higher threshold for inclusion, meaning that it selects less random effect factors in the model on average. This can be seen by noting the lower sensitivities and higher specificities in Figure~\ref{fig:M2_ROC} or Table~\ref{tab:M2_Selection}.

The reason for the difference between nWAIC and biWAIC in terms of the average number of random effect factors included is a result of the distribution from which they measure the sample means and variances needed to calculate the criterion. nWAIC, \eqref{eq:M2_nWAIC}, takes its sample means and variances based on only the distribution of $\bfy$, \eqref{eq:like_lat1}, the distribution which contains the random effects specification. biWAIC, \eqref{eq:M2_biWAIC}, however takes its sample means and variances from the marginalised distribution of $\bfy$ where $\boldsymbol\mu_{\textbf{y}}$ has been integrated out as detailed in Section~\ref{sec:M2_biWAIC}. As a result, like Bayesian 10-fold iCV, biWAIC takes into account the model fit of both $\bfy$ and $\boldsymbol\mu_{\textbf{y}}$.

Taking into account both of the probability distributions associated with the latent variables, \eqref{eq:like_lat1} and \eqref{eq:like_lat2}, better assesses the fit of the model and prevents the overfitting of the first distribution, \eqref{eq:like_lat1}, keeping the number of included random effect groups as a realistic level. nWAIC does not take into account how well the fixed effects, $\bfwg$, are able to predict the unknown true HI assays of a given pairs of reference and test strains, $\boldsymbol\mu_{\textbf{y}}$. nWAIC therefore picks the model which maximises the fit of \eqref{eq:like_lat1} regardless of the fit of \eqref{eq:like_lat2}, leading to the overfitting of \eqref{eq:like_lat1} and a potentially unrealistically high number of random effect groups. It is interesting however that we do not see a similar threshold with Bayesian 10-fold iCV, which also takes into account both parts of the latent variable likelihood. This is a consequence of the different sensitivity versus specificity trade offs given by criteria based on WAIC and those based on CV.

\section{Results for the A(H1N1) Dataset}
\label{sec:H1N1_results}

\begin{figure}[t]
				\centering
        \begin{subfigure}[b]{0.47\textwidth}
                \centerline{\includegraphics[width=\columnwidth]{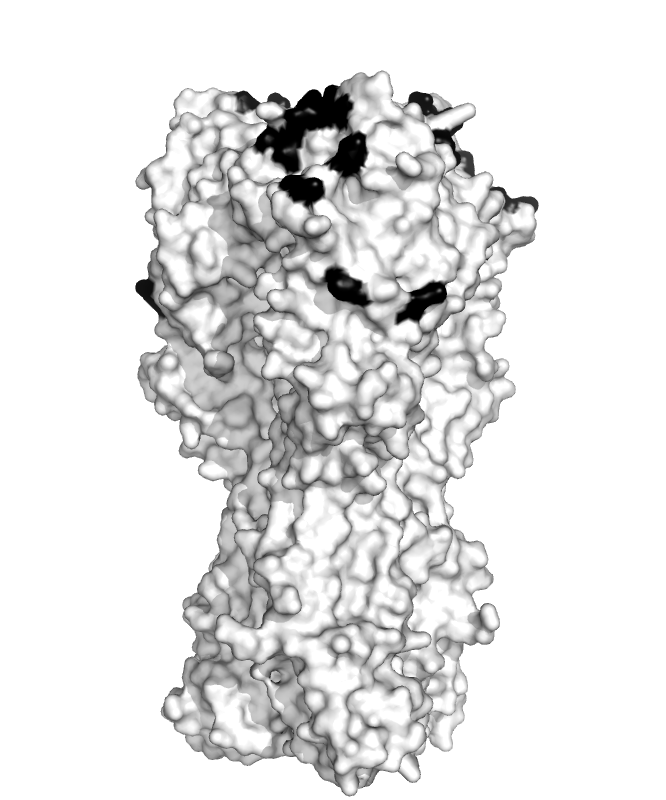}}
                \caption{Proven Residues}
                \label{fig:CapsidProven}
        \end{subfigure}
		\begin{subfigure}[b]{0.47\textwidth}
                \centerline{\includegraphics[width=\columnwidth]{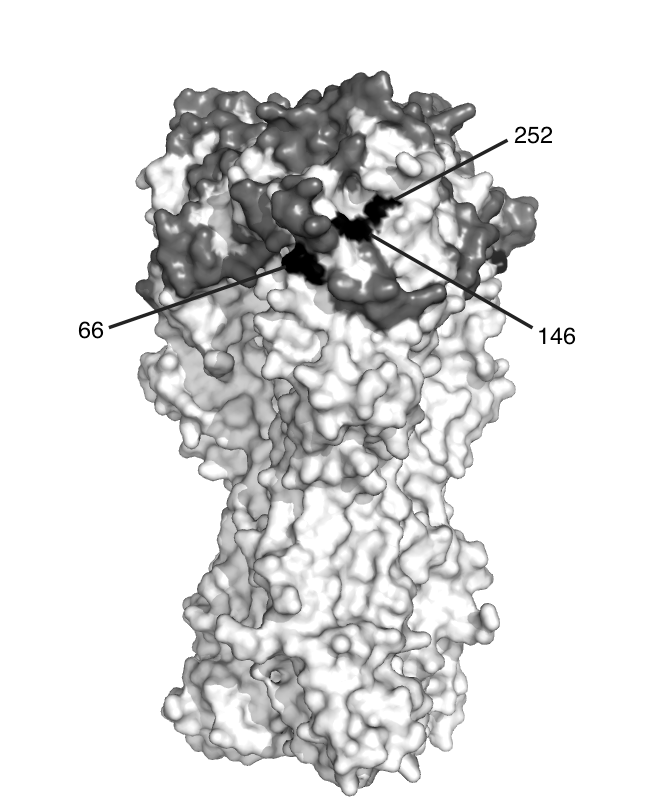}}
                \caption{Plausible Residues}
                \label{fig:CapsidPlausible}
        \end{subfigure}
	 \caption{Three-dimensional structure of the influenza A(H1N1) haemagglutinin protein showing the positions of proven and plausible antigenic residues identified using the eSABRE model. (a) Proven residues (black) selected by the eSABRE model. (b) Labelled plausible residues (black) where the biologically proven sites from Figure~\ref{fig:CapsidClassification} are shown in dark grey. The representation of the surface of HA is based on the resolved structure of influenza A(H1N1) strain A/Puerto Rico/8/34 \citep{Gamblin04}.}
				\label{fig:CapsidResults}
\end{figure}

We have applied the eSABRE model to the influenza A(H1N1) dataset described in Section~\ref{sec:H1N1} using the 8 possible combinations of random effect components. The biWAIC score, Section~\ref{sec:M2_biWAIC}, was then calculated for each of the models, with the model with the best biWAIC score including all random effect components. biWAIC was chosen to select the best model based on being more computationally feasible then 10-fold iCV and the results of Table~\ref{tab:M2_Selection} and Figures~\ref{fig:M2_Barplot} and~\ref{fig:M2_ROC}.

Having selected the model with the best selection of random effect components according to biWAIC, we have then compared the results in terms of variable selection to those achieved by \cite{Harvey16}. Using the eSABRE model as an exploratory tool, where we chose a relatively low threshold by taking the top $\hat{\pi} J$ variables with the highest marginal inclusion probabilities, we have selected 11 proven, 3 plausible and 3 implausible residues based on the classifications discussed in Section~\ref{sec:H1N1}. However for a fair comparison with the results of \cite{Harvey16}, we have also used a more conservative cut-off of, only selecting variables with a marginal inclusion probability of greater than 0.5, which gives the same false positive rate. Here we have selected 6 proven, 2 plausible and 1 implausible, compared to 4 proven, 0 plausible and 1 implausible selected in \cite{Harvey16}. These results show a clear improvement for the eSABRE model over the models used in \cite{Harvey16}.

Of the 11 proven residues selected and shown in Figure~\ref{fig:CapsidProven}, the eSABRE model, taking the top $\hat{\pi} J$ variables with the highest marginal inclusion probabilities, has identified one residue of the receptor-binding site, residue 187. Residue 187 is also part of the Sb antigenic site and we have also identified several other nearby residues (184, 189, 190 and 193) belonging to the same antigenic site. We have also identified proven residues from each of the other known H1 antigenic sites; Ca (141 and 142), Cb (69 and 74) and Sa (153). Finally we have identified the residue at position 130 to be important, a proven antigenic residue outside of these antigenic sites. An amino acid deletion at this site has been determined to have altered the structure of protein, causing a major change in the antigenic properties of the virus which required an update to the vaccine in the 1990s \citep{McDonald07}.

The 3 plausible residues identified (66, 146 and 252) can be found between the Ca and Cb antigenic sites as can be seen in Figure~\ref{fig:CapsidPlausible}. Figure~\ref{fig:CapsidPlausible} shows the proven antigenic sites of the A(H1N1) virus (dark grey) and the locations of the plausibles residues (black) between them. While all of the plausible residues could be plausibly antigenic, the most likely to be significant are 66 and 146 which both contact known antigenic sites on the surface of the protein. Residue 66 occurs directly between and immediately neighbouring two sections of the Ca and Cb antigenic sites. Both 146 and 252 also occur between sections of the Ca and Cb antigenic regions, however 146 is in a small indent on the surface and so while remaining very plausible, is perhaps less likely than 66. 252 is further away from the antigenic sites while still remaining in the head region of the virus close to known antigenic sites and was also picked up by \cite{Harvey16}.

Of the 3 implausible residues identified (43, 310 and 313), residue 43 has a reasonable explanation as to why it has been selected. Substitutions to the residue at position 43 are known to have occurred at the same branch of the phylogenetic tree as the important deletion referred to above at position 130, the proven antigenic residue selected by the eSABRE model described by \cite{McDonald07}, and in a second branch of the tree associated with a vaccine update. The by-chance co-occurrence of substitutions at 43 and genuine antigenic changes at multiple instances in the evolution of the virus provide an explanation for why 43 was identified both here and using other methods \citep{Harvey16}. The other two implausible residues identified, 310 and 313, are part of the stalk domain of the H1N1 virus and are unlikely to have a significant antigenic effect. It is however noteworthy that 313 was identified at a later, non-comparable stage of the analysis of \cite{Harvey16} which involved the identification of specific amino acid substitutions that correlated with points in the evolution of the virus where the antigenicity changed.

\section{Conclusions and Future Work}

We have developed a novel hierarchical Bayesian model, called eSABRE, for detecting antigenic sites in virus evolution, with particular focus on the influenza virus. Our model is based on a predecessor, called SABRE, that was developed in the context of studying antigenicity in the foot-and-mouth-disease virus. However, the SABRE model turned out to be computationally too inefficient for larger data sets, as typically available for the influenza virus. We have demonstrated that by building a new structure of the hierarchical model, we can not only improve the computational efficiency by several orders of magnitude, but we also significantly improve the prediction accuracy by making the model more consistent with the format of the data (see Tables~\ref{tab:M2_AUROC} and~\ref{tab:M2_AUROC_FMDV}). In addition to testing the eSABRE model, we have also looked at the best way of selecting random effects coefficients. In Section~\ref{sec:M2_biWAIC} we proposed biWAIC as a new method for selecting random effect components in the latent variable models with the structure of the eSABRE model, where it is not possible to apply iWAIC. The results of Table~\ref{tab:M2_Selection} and Figures~\ref{fig:M2_Barplot} and~\ref{fig:M2_ROC} show how biWAIC properly accounts for the distribution of the latent variables, resulting in a more realistic number of random effect components being included compared to nWAIC and a smaller computational cost than Bayesian iCV.

Section~\ref{sec:H1N1_results} demonstrates how the eSABRE model, together with biWAIC, can be effectively applied to a large real life Influenza datasets. In Section~\ref{sec:H1N1_results} we show how the improvement in computational efficiency demonstrated in Table~\ref{tab:M2_AUROC} allows us to make use of the full H1N1 dataset rather than a reduced version as was required for the conjugate SABRE model in \cite{Davies17_COST}. The results from using the full H1N1 dataset and properly accounting for the error in the data collection process through the eSABRE model show an improvement in the selection of antigenic variables in the H1N1 datasets.

Further work involves investigating how different types of amino acid change at the same residue position on the structure affect antigenic variability. The data currently consists of indicators of amino acid change that are identical regardless of which particular amino acids are involved. However, given the range of biophysical properties among different amino acids, we expect the antigenic impact of a change to depend on both the location on the structure and the particular amino acids involved. In \cite{Reeve16} and \cite{Harvey16} variables were used that indicated a particular amino acid change at a given location. This will significantly increase the number of variables that must be selected from and therefore it is likely that the model will require additional information in order to prevent spurious results. Latent Gaussian processes can be used to include the additional information, e.g. \cite{Flippone13}, and will allow us to account for (a) differences in the antigenic impact of amino acid substitutions that depend on the amino acids involved and (b) similarities between changes of a certain type that occur at the same, or similar, locations on the protein surface.

\begin{algorithm}[b]
\SetAlgoLined
\SetKwInOut{Input}{Input}
\SetKwInOut{Output}{Output}
 \Input{Nominal hyperparameters $\hyperpt$}
 \Output{Corrected hyperparameters $\hyperp$}

 \While{prior $p(\param|\hyperp)$ is inconsistent with actual prior knowledge}{
  Chose a new nominal hyperparameter vector $\hyperpt$\;
  Run a multiple partition MCMC simulation without data\;
  Compute the Kullback-Leibler divergence by application of (\ref{eq:KLE})\;
  Obtain new corrected hyperparameters $\hyperp$ by application of (\ref{eq:KLmin})\;
 }
 \caption{Prior correction for combining sub-posteriors \label{algo:one}}
\end{algorithm}

If the data set $\Data$ is so large that the MCMC run time becomes onerous, the divide-and-conquer strategy of \cite{NemethSherlock17} can be applied, which is based on partitioning the data, $\Data = \{\Data_1,\Data_2,\ldots\Data_{\Npart}\}$, running $\Npart$ separate MCMC simulations in parallel to obtain the sub-posterior  
distributions $p(\param|\Data_i)$, $i=1,\ldots,\Npart$, 
and then combining these sub-posteriors along the following lines:
\begin{equation}
\label{eq:Pmcmc}
p(\param|\Data,\hyperp) \propto p(\Data|\param)p(\param|\hyperp) 
\; = \;
\prod_{i=1}^{\Npart} \Big(p(\Data_i|\param) \pap(\param|\hyperp)\Big)
\;\propto \;
\prod_{i=1}^K p(\param|\Data_i,\hyperp)
\end{equation}
where
\begin{equation}
\pap(\param|\hyperp) \; \propto \; 
\sqrt[\Npart]{p(\param|\hyperp)}
\end{equation}
 The practical difficulty in applying this scheme to a complex hierarchical Bayesian model, like the eSABRE model, is 
 the intractability of the new prior distribution $\pap(\param|\hyperp)$. In practice, all we can do is set $\pap(\param|\hyperp)=p(\param|\hyperp)$, i.e.
 we effectively introduce a modified prior distribution
 \begin{equation}
q(\param|\hyperpt) \; \propto \; 
[p(\param|\hyperpt)]^{\Npart}
\end{equation}
The objective, then, is to find hyperparameters $\hyperpt$ so that $q(\param|\hyperpt)$ matches as closely as possible the original prior 
distribution $p(\param|\hyperp)$, e.g. by minimising the Kullback-Leibler divergence
\begin{equation}
\KLE(\hyperp,\hyperpt)
\; = \; 
\int q(\param|\hyperpt) \log\left(\frac{q(\param|\hyperpt)}{p(\param|\hyperp)} \right) d\param
\end{equation}
This is infeasible, due to the intractability of  $E(\hyperp,\hyperpt)$ as a function of $\hyperpt$. However, what we can do instead is minimise 
$\KLE(\hyperp,\hyperpt)$ for fixed $\hyperpt$ with respect to $\hyperp$, since
\begin{eqnarray}
\KLE(\hyperp,\hyperpt)
& = &
- \int q(\param|\hyperpt) \log\Big(p(\param|\hyperp)\Big) d\param
\,+\, C_{\hyperpt}
\\
& \approx & 
- \sum_{i=1}^N
\log\Big(p(\param_i|\hyperp)\Big)
\,+\, C_{\hyperpt}
\label{eq:KLE}
\end{eqnarray}
where $C_{\hyperpt}$ is a constant with respect to $\hyperp$, and 
$\{\param_i\}_{i=1}^{N}$ is a sample from the distribution $q(\param|\hyperpt)$, which is 
straightforwardly obtained by running the parallel multi-partition MCMC algorithm of \cite{NemethSherlock17} without data.
So while we cannot specify the modified prior distribution $q(\param|\hyperpt)$ directly, we can find the original prior distribution $p(\param|\hyperp)$
that provides the closest match:
\begin{equation}
\mathrm{argmin}_{\hyperp} \KLE(\hyperp,\hyperpt)
\label{eq:KLmin}
\end{equation}
where the minimisation can be achieved with standard iterative optimisation algorithms.  
If that prior distribution is unacceptable, due to insufficient agreement with the actual prior knowledge, a new hyperparameter vector $\hyperpt$ can be tried until the agreement  is acceptable.   A pseudo code representation of the algorithm is shown in Algorithm~\ref{algo:one}.
Note that MCMC simulations without data correspond to the extreme annealing end of a parallel tempering scheme, where convergence and mixing are usually fast, and the computational costs are thus comparatively low. Also, the prior distribution is usually chosen to be vague due to lack of actual prior knowledge. In that case, the condition in the {\tt While} statement on line 1 of the algorithm can be replaced by the following less stringent criterion: {\tt While $p(\param|\hyperp)$ is insufficiently disperse}.

\begin{figure}[tb]
\centering
\includegraphics[scale=0.3]{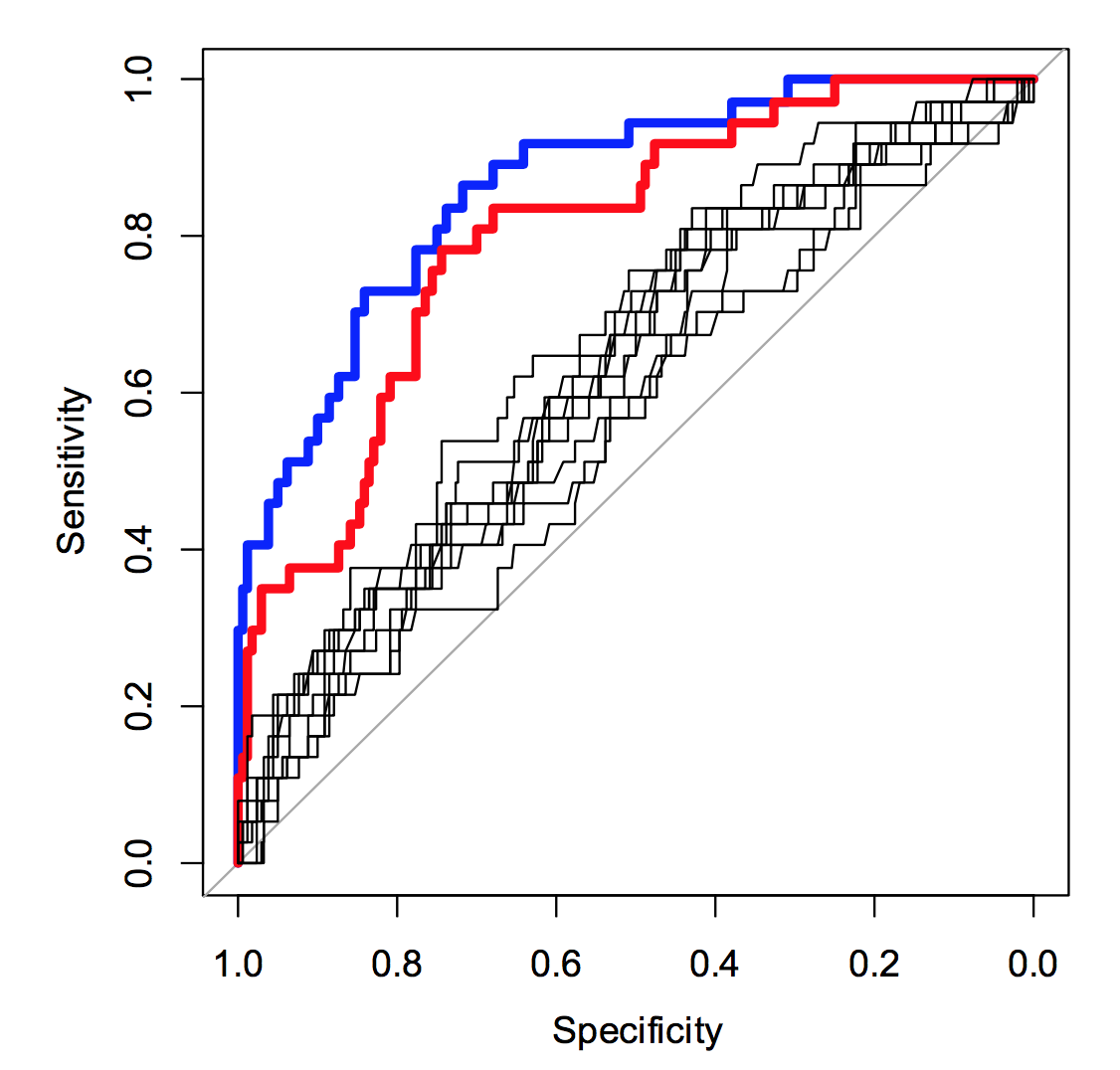}
\includegraphics[scale=0.3]{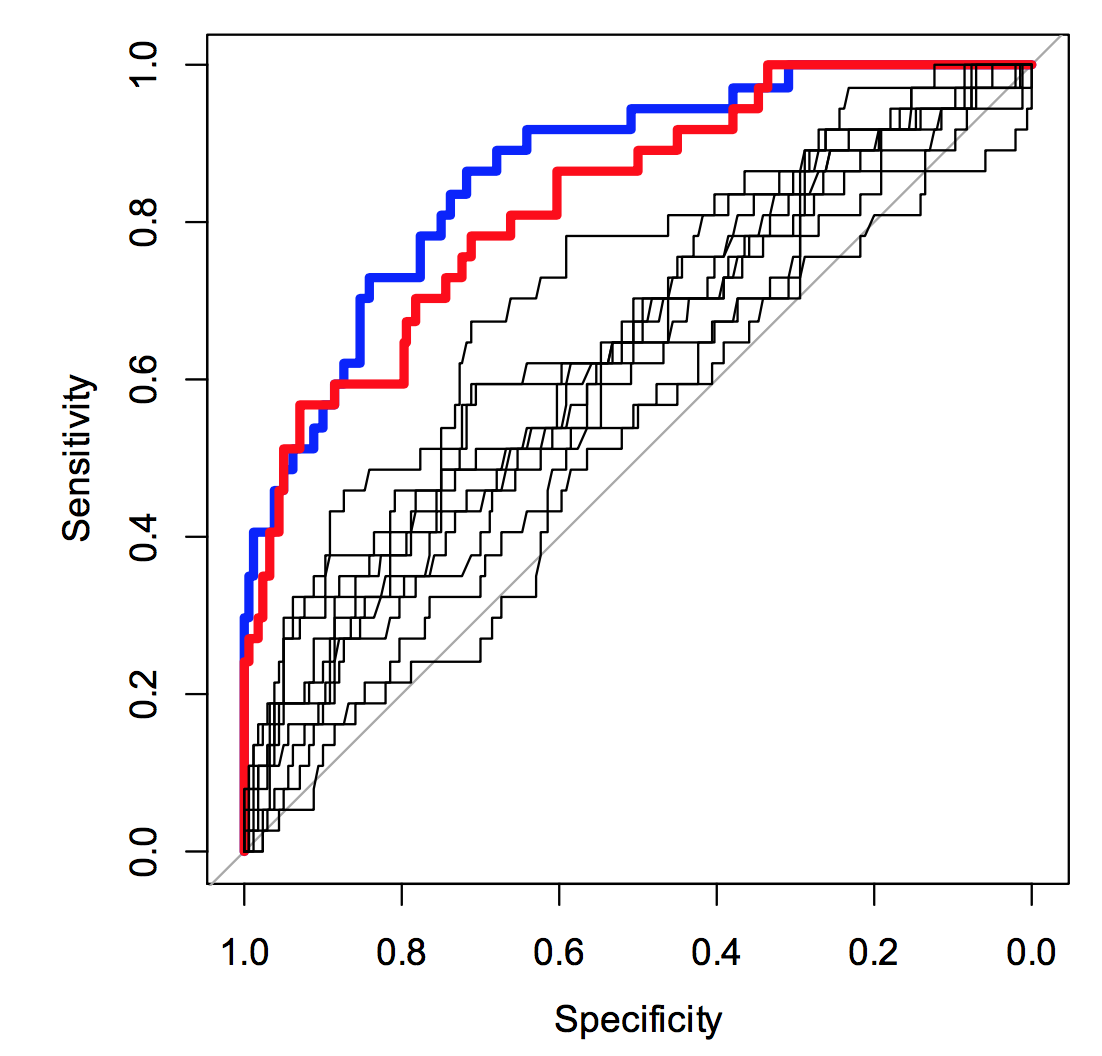}
\caption
{
\label{fig:subMCMC}
{\bf ROC curves for assessment of the sub-posterior method.}
The blue lines show the ROC curves for the proper posterior distribution of variable selection, obtained by running an MCMC simulation on the whole data set. The thin  black lines show the ROC curves of the sub-posterior distributions, obtained by running ten MCMC simulations in parallel on ten non-overlapping subsets of the data, each containing a tenth of the total data. The red lines show the ROC curves obtained by combining the posterior distributions according to (\ref{eq:Pmcmc}). Left panel: no adjustment of the hyperparameters. Right: adjustment of the hyperparameters, as discussed in the text. The improvement in the agreement between the two curves is clearly noticeable. The simulations were carried out by Costanza Calzolari as part of her final year Honours project at the University of Glasgow, supervised by the authors.
}
\end{figure}

We have carried out a simple proof of concept study on simulated data, similar to those described in Section~6, where we have compared the performance of the sub-posterior method with standard MCMC. The data were partitioned into 10 non-overlapping subsets, each containing a tenth of the original data. Ten MCMC simulations were run in parallel, and the sub-posteriors were then combined, based on (\ref{eq:Pmcmc}), as described in \cite{NemethSherlock17}. These results were compared with a standard MCMC simulation on the complete data set. The results are shown in Figure~\ref{fig:subMCMC} and suggest that when correcting for the mismatch between the hyperparameters, a similar performance can be achieved as with standard MCMC, but at substantially reduced computational costs.

A more extensive exploration is beyond the remit of the present paper, but will provide the prospect of making the proposed eSABRE model applicable to data from future virus sequencing projects that are much bigger than those included in the present study.


\bibliographystyle{chicago}
\bibliography{References}

\end{document}